\documentclass[a4paper,11pt]{article}
\pdfoutput=1 

\usepackage{jcappub} 

\usepackage[T1]{fontenc} 
\usepackage{units}
\usepackage{float}
\usepackage{lineno}
\usepackage{amsmath}
\usepackage{booktabs}
\usepackage{enumitem}
\usepackage{caption}
\usepackage{subcaption}
\usepackage{glossaries}
\usepackage{csquotes}

\newcounter{raffil} 
\newcommand{\Affiliation}[2]{%
    \affiliation[\refstepcounter{raffil}\label{aff:#1}\ref{aff:#1}]{#2}%
}

\newcounter{raddr} 
\newcommand{\Address}[2]{%
    \affiliation[\refstepcounter{raddr}\label{aff:#1}\ref{aff:#1}]{#2}%
}

\NewDocumentCommand{\Author}{ooom}{%
    \author[\ref{aff:#1}\IfValueT{#2}{,\ref{aff:#2}}\IfValueT{#3}{,\ref{aff:#3}}]{#4}
}

\title{\boldmath Modeling the Effect of the Heliospheric Magnetic Field on Cosmic Ray Muon Shadows}

\begin{document}


\Author[FNAL]{P.~Adamson}

\Author[Iowa]{I.~Anghel}

\Author[Cincinnati]{A.~Aurisano}

\Author[Oxford]{G.~Barr}

\Author[Cambridge][Lancaster]{A.~Blake}

\Author[Texas]{S.~V.~Cao}

\Author[Texas][Wisconsin]{T.~J.~Carroll}

\Author[UFG]{C.~M.~Castromonte}

\Author[Manchester]{R.~Chen}

\Author[FNAL]{S.~Childress}

\Author[Tufts]{J.~A.~B.~Coelho}

\Author[Texas]{S.~De~Rijck}

\Author[Manchester]{J.~J.~Evans}

\Author[Harvard]{G.~J.~Feldman}

\Author[DallasU]{W.~Flanagan}

\Author[Duluth][CSU]{S.~Fogarty}

\Author[Minnesota]{M.~Gabrielyan}

\Author[Tufts]{H.~R.~Gallagher}

\Author[UCL]{S.~Germani}

\Author[UFG]{R.~A.~Gomes}

\Author[ANL]{M.~C.~Goodman}

\Author[USP]{P.~Gouffon}

\Author[Pitt]{N.~Graf}

\Author[FNAL]{N.~Grossman}

\Author[Warsaw]{K.~Grzelak}

\Author[Duluth]{A.~Habig}

\Author[FNAL]{S.~R.~Hahn}

\Author[Sussex]{J.~Hartnell}

\Author[FNAL]{R.~Hatcher}

\Author[UCL]{A.~Holin}

\Author[Texas]{J.~Huang}

\Author[Houston]{L.~W.~Koerner}

\Author[WandM]{M.~Kordosky}

\Author[FNAL]{A.~Kreymer}

\Author[Duluth]{J.~Krueger}

\Author[Texas]{K.~Lang}

\Author[FNAL]{P.~Lucas}

\Author[Tufts]{W.~A.~Mann}

\Author[Minnesota][deceased]{M.~L.~Marshak}

\Author[Tufts]{N.~Mayer}

\Author[Texas]{R.~Mehdiyev}

\Author[Minnesota]{J.~Meier}

\Author[Minnesota]{W.~H.~Miller}

\Author[LANL][deceased]{G.~Mills}

\Author[Pitt]{D.~Naples}

\Author[WandM]{J.~K.~Nelson}

\Author[UCL]{R.~Nichol}

\Author[UCL]{J.~O'Connor}

\Author[FNAL]{R.~B.~Pahlka}

\Author[FNAL]{\v{Z}.~Pavlovi\'{c}}

\Author[Minnesota]{G.~Pawloski}

\Author[UCL]{A.~Perch}

\Author[UCL]{M.~M.~Pf\"{u}tzner}

\Author[Texas]{D.~D.~Phan}

\Author[FNAL]{R.~K.~Plunkett}

\Author[FNAL]{N.~Poonthottathil}

\Author[Stanford]{X.~Qiu}

\Author[WandM]{A.~Radovic}

\Author[Texas]{P.~Sail}

\Author[Iowa][ANL]{M.~C.~Sanchez}

\Author[Texas]{A.~Schreckenberger}

\Author[FNAL]{R.~Sharma}

\Author[Duluth][Corresponding]{N.~Skuza}

\Author[Cincinnati]{A.~Sousa}

\Author[Otterbein]{N.~Tagg}

\Author[UCL]{J.~Thomas}

\Author[Cambridge]{M.~A.~Thomson}

\Author[Manchester]{A.~Timmons}

\Author[Cincinnati]{J.~Todd}

\Author[UFG][ORNL]{S.~C.~Tognini}

\Author[Harvard]{R.~Toner}

\Author[FNAL]{D.~Torretta}

\Author[WandM]{P.~Vahle}

\Author[Oxford][RAL]{A.~Weber}

\Author[UCL]{L.~H.~Whitehead}

\Author[Stanford][deceased]{S.~G.~Wojcicki}

\collaboration{The MINOS+ Collaboration}

\Affiliation{ANL}{Argonne National Laboratory, Argonne, Illinois 60439, USA}
\Affiliation{Cambridge}{Cavendish Laboratory, University of Cambridge, Cambridge CB3 0HE, United Kingdom}
\Affiliation{Cincinnati}{Department of Physics, University of Cincinnati, Cincinnati, Ohio 45221, USA}
\Affiliation{CSU}{Department of Physics, Colorado State University, Fort Collins, CO 80523-1875, USA}
\Affiliation{DallasU}{University of Dallas, 1845 E Northgate Drive, Irving, Texas 75062 USA}
\Affiliation{FNAL}{Fermi National Accelerator Laboratory, Batavia, Illinois 60510, USA}
\Affiliation{UFG}{Instituto de F\'{i}sica, Universidade Federal de Goi\'{a}s, Goi\^{a}nia, Goi\'{a}s, 74690-900, Brazil}
\Affiliation{Harvard}{Department of Physics, Harvard University, Cambridge, Massachusetts 02138, USA}
\Affiliation{Houston}{Department of Physics, University of Houston, Houston, Texas 77204, USA}
\Affiliation{Iowa}{Department of Physics and Astronomy, Iowa State University, Ames, Iowa 50011, USA}
\Affiliation{Lancaster}{Lancaster University, Lancaster, LA1 4YB, UK}
\Affiliation{LANL}{Los Alamos National Laboratory, Los Alamos, New Mexico 87545, USA}
\Affiliation{Manchester}{Department of Physics and Astronomy, University of Manchester, Manchester M13 9PL, United Kingdom}
\Affiliation{Duluth}{Department of Physics and Astronomy, University of Minnesota Duluth, Duluth, Minnesota 55812, USA}
\Affiliation{Minnesota}{School of Physics and Astronomy, University of Minnesota Twin Cities, Minneapolis, Minnesota 55455, USA}
\Affiliation{ORNL}{Oak Ridge National Laboratory, Oak Ridge, TN, USA}
\Affiliation{Otterbein}{Otterbein University, Westerville, Ohio 43081, USA}
\Affiliation{Oxford}{Subdepartment of Particle Physics, University of Oxford, Oxford OX1 3RH, United Kingdom}
\Affiliation{Pitt}{Department of Physics, University of Pittsburgh, Pittsburgh, Pennsylvania 15260, USA}
\Affiliation{RAL}{Rutherford Appleton Laboratory, Science and Technology Facilities Council, Didcot, OX11 0QX, United Kingdom}
\Affiliation{USP}{Instituto de F\'{i}sica, Universidade de S\~{a}o Paulo,  CP 66318, 05315-970, S\~{a}o Paulo, SP, Brazil}
\Affiliation{Stanford}{Department of Physics, Stanford University, Stanford, California 94305, USA}
\Affiliation{Sussex}{Department of Physics and Astronomy, University of Sussex, Falmer, Brighton BN1 9QH, United Kingdom}
\Affiliation{Texas}{Department of Physics, University of Texas at Austin, Austin, Texas 78712, USA}
\Affiliation{Tufts}{Department of Physics and Astronomy, Tufts University, Medford, Massachusetts 02155, USA}
\Affiliation{UCL}{Physics and Astronomy Dept., University College London, Gower Street, London WC1E 6BT, United Kingdom}
\Affiliation{Warsaw}{Department of Physics, University of Warsaw, PL-02-093 Warsaw, Poland}
\Affiliation{WandM}{Department of Physics, William \& Mary, Williamsburg, Virginia 23187, USA}
\Affiliation{Wisconsin}{Department of Physics, University of Wisconsin-Madison, Madison, Wisconsin 53706, USA}
\Affiliation{deceased}{Deceased.}
\Address{Corresponding}{Corresponding Author: nicholasskuza@gmail.com}

\begin{abstract} {Shadows cast in the cosmic ray (CR) muon sky by the Sun were located using muon data from the MINOS far detector in Northern Minnesota. The shadows were observed independently across three time periods; near solar minimum, near solar maximum, and over the entire 13 year span of the data. A distribution of muon positions for each shadow was then sampled to simulate CR motions near the Sun using the Parker spiral model of the Heliospheric Magnetic Field (HMF) and a dipole model of the Geomagnetic Field (GMF). The resulting particle distributions were then compared to their position with respect to the Sun. Results show that the Parker spiral model is most consistent with the solar minimum shadow and least consistent with the solar maximum shadow, as expected. The simple Parker spiral is more consistent with
the data for a harder CR spectrum than is actually present, indicating the need for a more detailed HMF model. Plausible modifications to the Parker spiral model which would affect the overall shift of the Sun's CR shadow are discussed.}

\end{abstract}
\maketitle

\flushbottom
\newpage
\section{Introduction}
\label{sec:intro}

\subsection{Cosmic Rays and Cosmic Ray Muons}

Cosmic rays (CRs) are relativistic charged particles which emanate from multiple extraterrestrial sources. They are composed of protons ($\approx \unit[87]{\%}$), alpha particles ($\approx \unit[12]{\%}$) and heavier ions ($\approx \unit[1]{\%}$) \cite{Travnicek:2004PHD}. Their energies follow a power law distribution of the form $\propto E^{\gamma}$ where $\gamma$ is negative in most regimes. Lower energy CRs are more likely to come from sources within the solar system such as the Sun, while Galactic Cosmic Rays (GCRs) are generally higher energy and originate from outside the solar system \cite{Adamson:ApP}. Like all charged particles, CRs and GCRs are influenced by electromagnetic fields. In our solar system the dominant electromagnetic fields are the heliospheric magnetic field (HMF) and the planetary fields. Once GCRs strike the upper atmosphere of the Earth, with a flux of approximately $\unit[1000]{m^{-2} s^{-1}} $, they produce a shower of secondary particles. Among these secondary particles are muons, charged particles which are prime candidates to penetrate rock layers and interact with underground detectors \cite{Gaisser:CR}. The higher the energy of the primary CR, the more likely the secondary muon is to reach an underground detector \cite{Adamson:ApP}. 

\subsection{MINOS Far Detector}
The Main Injector Neutrino Oscillator Search (MINOS) far detector is a magnetized scintillator and steel tracking calorimeter located underground at a depth of 2070 meters water equivalent (mwe) in the Soudan Mine (47° 49’ 13.3” N, 92° 14’ 28.5” W) in Northern Minnesota. Its primary purpose was to detect neutrinos, but its great depth and wide acceptance combined with the flat overburden of the Soudan site allows it to function as a CR muon detector \cite{FD:description}. The \unit[5.4]{kton} detector is composed of 486 octagonal planes, each \unit[8]{m} wide consisting of a \unit[2.54]{cm} thick steel plate and a \unit[1.0]{cm} thick scintillator plane. Adjacent planes are separated by a \unit[2.4]{cm} air gap. It is \unit[30]{m} long and has a total aperture of $\unit[6.91 \times 10^{6}]{cm^{2}sr}$  \cite{Adamson:ApP} for the GCRs selected in this analysis \cite{Fogarty:2019DDC}. The detector observes muons with a minimum surface energy of \unit[0.7]{TeV} and a mean value of approximately \unit[1.0]{TeV} \cite{muonEnergyDistribution}.

The MINOS far detector detected more than 10\textsuperscript{8} candidate muons over a 13 year period from July 2003 to July 2016. With knowledge of the angles of incidence and timing of these detections, the trajectories of GCRs can be mapped back onto the sky \cite{Fogarty:2019DDC}. With only a few months of data it is possible to locate features of interest in an otherwise uniform distribution of GCRs \cite{Adamson:ApP}. These features of interest are caused by objects within the solar system impeding incident GCRs en route to Earth. Objects such as the Sun and the Moon subtend a large enough portion of the sky to produce observable shadows. Once located and characterized, the Sun shadow can be used to probe the structure of the HMF. 

\subsection{Solar and Lunar Muon Shadows}
Most of the muons observed underground in the \unit[1.0]{TeV} energy regime originate from particle showers caused by GCRs striking the upper atmosphere. These muon energies correspond to a primary CR energy of approximately \unit[10]{TeV} \cite{Adamson:ApP}. Using the muon data from the far detector, typical trajectories of incident GCRs can be reconstructed to look for features in the CR distribution. Shadows that subtend an angle  of $\unit[0.52]{^\circ}$ are produced by the Sun and Moon absorbing incident GCRs. These shadows are characterized by regions of little to no particle density \cite{Adamson:ApP}. The shadows are not easily distinguished from the background because of muon deflections. There is also a shift produced by the HMF and Earth's geomagnetic field (GMF). 

The GMF is well-mapped and understood, however the much larger HMF is complex and difficult to model. \cite{Kallenrode:2004plasma}. Using the location of the Sun’s muon shadow, its measured shift, and the reconstructed trajectories of primary GCRs, the behavior of the HMF can be inferred using Monte Carlo simulations. The goal of these simulations is to determine whether the general behavior of the shadows is consistent with theory and, by testing differing models of the HMF, to determine which version of the magnetic field is most consistent with data.

\section{Data and Preliminary Analysis}
\label{sec:data&analysis}

\subsection{Data Selection and Time Frames}

Only single muon events are used to infer the GCR shadows. In some instances, higher energy GCRs will produce more than one muon visible in the detector, but these events only make up $3.1\%$ of all events. The exclusion of multiple muon events is done for the convenience of having only one apparent GCR trajectory per event, however multiple muon events are useful for the purpose of characterizing muon deflections in the rock overburden that are incurred by single-muon trajectories \cite{Fogarty:2019DDC}.  Muons must intercept enough planes and travel a significant distance within the detector to reconstruct their trajectories with confidence. For this analysis, only particles that travel at least \unit[1.55]{m} within the detector and intercept at least 9 of the 486 planes that make up the detector were selected. Only particles with incident zenith angles less than 80\textdegree\ were kept to ensure that only muons sourced from the sky were included. Some interaction vertices exist outside the detector due to badly reconstructed trajectories; these particles have also been excluded. Upon applying these criteria, $84.8\%$ of the $\unit[1.3987 \times 10\textsuperscript{8}]{muons}$ initially selected, are retained for analysis. 

The Sun's magnetic field and therefore the HMF vary throughout the 11-year solar cycle \cite{noaaSolarCycle}. Periods known as solar minimum and solar maximum are included within the data set. The time frame of the data allows for the study of the differences between the solar CR shadows during different magnetic conditions near the Sun. The solar minimum data set contains $\unit[3.4888 \times 10\textsuperscript{7}]{muons}$ from January 2007 to December 2009. The solar maximum data set contains $\unit[2.8178 \times 10\textsuperscript{7}]{muons}$ from January 2012 to December 2014. 

\begin{figure} [htbp] 
\centering
    \centering
    \includegraphics[width=0.95\textwidth]{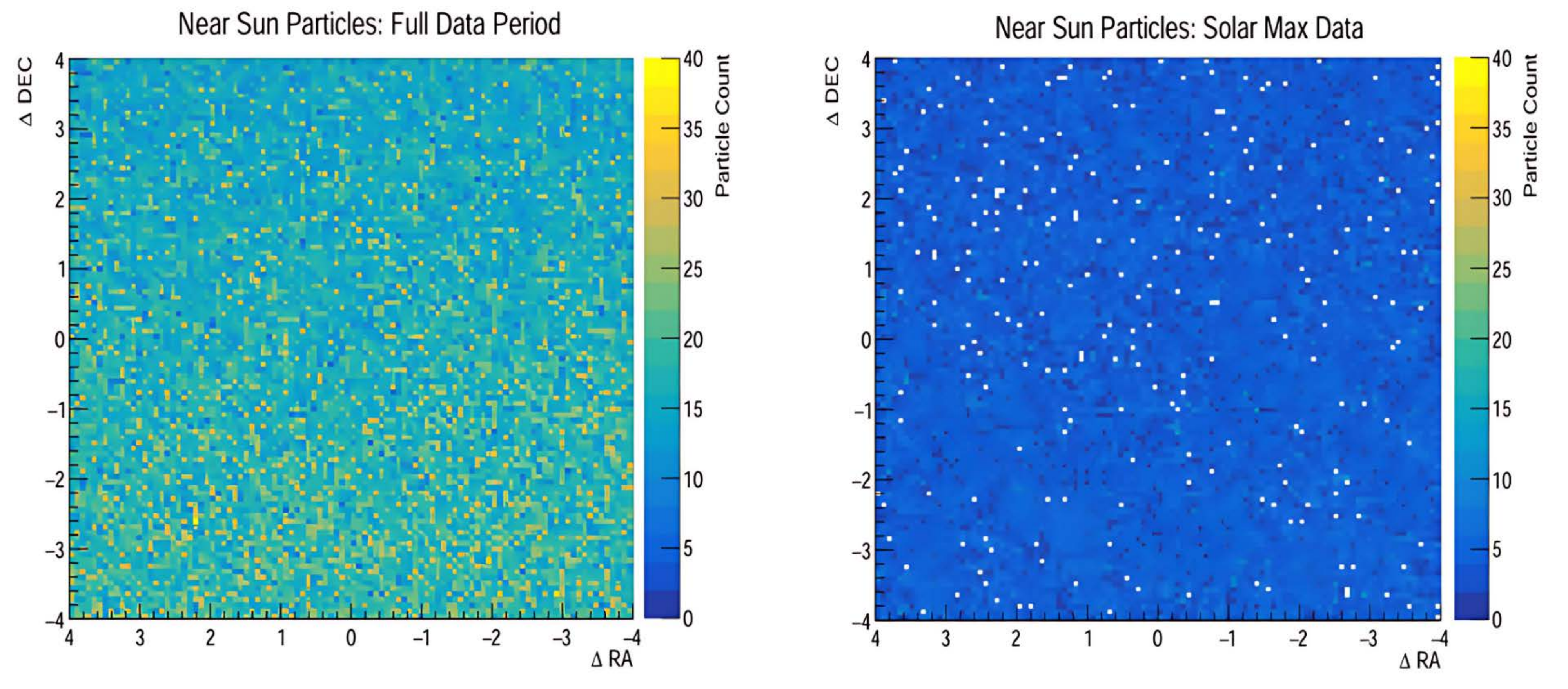}
    \label{fig:allData}
    
\caption[Data Histograms]{Particles detected near the Sun for the whole time range (left) and the 3 year period near solar max (right). The left plot has less than 20 particles per bin while the right plot has less than 5 particles per bin. Solar min data appears similar to the solar max data. $\Delta$RA and $\Delta$DEC refer to the angular distance in terms of right ascension ($\Delta$RA) and declination ($\Delta$DEC) on the sky relative to the Sun's known location in degrees. The Sun's shadow is not discernible in these particle count maps.}
\label{fig:allDataMaxData}
\end{figure}

These particle counts are across the whole sky, but only particles with reconstructed trajectories that pass near the Sun can be used to locate the Sun shadow. The histograms in Fig.~\ref{fig:allDataMaxData} show the distribution of near-Sun particles. Of all particles within $4^{\circ}$ of the Sun that were selected, there were fewer than 20 particles in most bins. This number is proportionally less for the solar maximum and solar minimum data because the time periods are only 3 years instead of 13. The Sun shadows are not immediately recognizable due to the CR muon scattering, as the distributions are generally flat. Statistical techniques must be used to locate and characterize the shadows.

\subsection{Sky Simulations and Shadow Templates}

The first step to understand the effect of the HMF is to locate the shadows. The shadow is hidden by several factors, but can still be found using a $\Lambda$ probability test. It is a log scale probability where positive values indicate that a certain collection of bins fits a theoretical shadow template better than a flat background of particles. The formula for such a test is shown in Eq.~\ref{logProbC}. This is a modified form of the same equation provided by \cite{Cash:1979}:

\begin{equation}
\centering
    \Lambda = \lambda(x\textsubscript{s}, y\textsubscript{s}, 0) - \lambda(x\textsubscript{s}, y\textsubscript{s}, I\textsubscript{s}) = 2\sum_{i=1}^{n\textsubscript{i}} [N\textsubscript{back} - N\textsubscript{th} + N\textsubscript{obs} \: \ln \frac{N\textsubscript{th}}{N\textsubscript{back}} ].
    \label{logProbC}
\end{equation} The value $\Lambda$ represents the likelihood that the shadow is located at a particular point. The $\lambda(x\textsubscript{s}, y\textsubscript{s}, 0)$ term refers to a source location on the sky $(x\textsubscript{s}, y\textsubscript{s})$ with no shadow present as indicated by the 3rd argument, 0. The $\lambda(x\textsubscript{s}, y\textsubscript{s}, I\textsubscript{s})$ term refers to the same location with a fractional decrease in the background intensity caused by the shadow, represented by $I\textsubscript{s}$. $I\textsubscript{s}$ is determined from a shadow template and $(x\textsubscript{s}, y\textsubscript{s})$ refers to a particular bin location in a 2D histogram. The larger the positive $\Lambda$ values, the more probable the pattern was produced from a shadow distribution rather than a flat one. A single $\Lambda$ value is generated for a particular central bin by summing over an entire shadow pattern centered on that bin. The observed bin counts, $N\textsubscript{obs}$, are compared to a set of bins that represent a theoretical shadow, $N\textsubscript{th}$, and a theoretical flat background distribution, $N\textsubscript{back}$. 

The background histograms should be as smooth as possible and sample the whole sky rather than simply particles near the Sun. To do this, a muon is generated with a random position (zenith, azimuth) using a Julian date (i.e., the date of ``observation'') from the data set to reflect actual live time distributions. Since the date and position are generated independently, any shadow features that may be present will not appear in the background histogram. The uncertainty on the number of particles in any particular bin is Poissonian, which implies that increasing the number of skies to average over reduces the fractional uncertainty.

The $N\textsubscript{th}$ values referenced in Eq.~\ref{logProbC} can be represented by a theoretical shadow template. It is assumed that the GCRs pass by the Sun and all particles that strike the Sun are absorbed. This would leave a $0.52^{\circ}$ hole from the perspective of an observer on Earth. The particles would then traverse the atmosphere and penetrate the \unit[1070]{mwe} of rock layers between the atmosphere and the detector. The template starts as a flat distribution of particles with a $0.52^{\circ}$ diameter hole. The observed scattering of di-muon events in the rock layers above the detector are then applied to those points to produce the template. Details concerning the template construction are described in Ref. \cite{Adamson:ApP}.

The template represents what the solar shadow should look like after it has penetrated the rock layers and reached the detector. The minimum of the Sun shadow template is $\approx \unit[80]{\%}$ of the background field, so a fractional uncertainty of significantly less than $\sigma\textsubscript{frac} = 0.2$ is required to provide statistically significant results for any particular bin $n\textsubscript{ij}$. A fractional uncertainty of $\sigma\textsubscript{frac} \approx 0.01$ requires the average value of $n\textsubscript{ij}$ be approximately $10^{4}$. This average bin count was selected as the goal for each background histogram. There were 600 skies of particles simulated for the full data set and 3000 skies worth for solar max/min. Once the histograms were produced, the bins were divided by the number of skies to determine the average bins per sky quantity throughout the background template.

\subsection{Locating the Shadows}

\begin{figure} [htbp] 
\centering
    \includegraphics[width=0.95\textwidth]{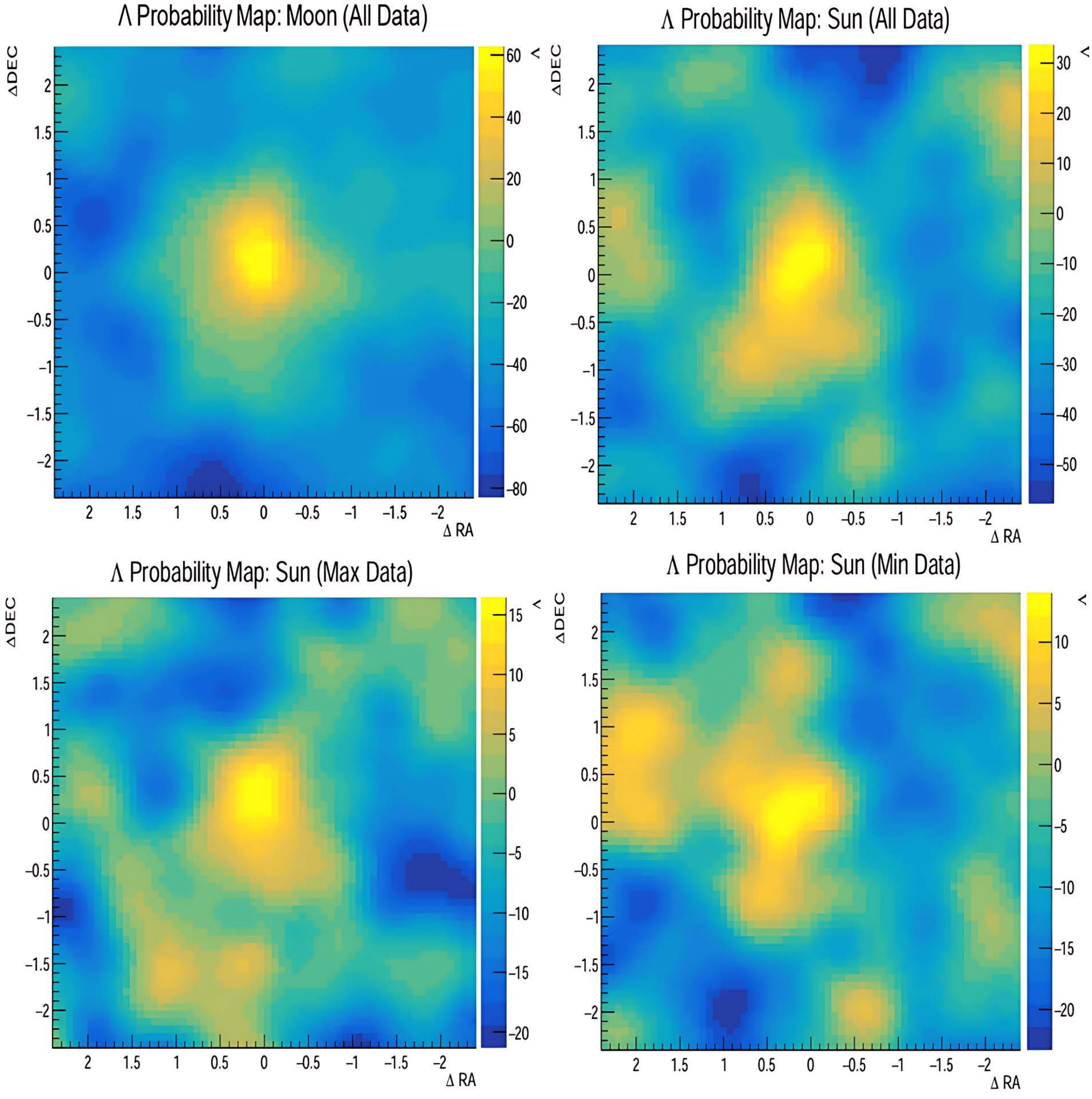}
\caption[Max-Likelihood Probability Histograms]{$\Lambda$ max-likelihood probability histograms for the Moon shadow (top left) and for the Sun shadow (top right); the Sun shadow during solar min (bottom right); and the Sun shadow during solar max (bottom left). Larger $\Lambda$ values correspond to locations where the CR muon shadow is likely to be. The maximum values ($\Lambda\textsubscript{max}$) are 62.76, 33.57, 14.01, and 16.44 going clockwise from the top left. Note that the $\Lambda$-scale for each histogram is different. The $\Lambda\textsubscript{max}$ locations relative to center are ($\Delta$RA, $\Delta$DEC) = $(0.04^{\circ}, 0.12^{\circ}); (0.12^{\circ}, 0.12^{\circ}); (0.04^{\circ}, 0.28^{\circ}); (0.28^{\circ}, 0.04^{\circ})$ once again clockwise from top left.}
\label{fig:lambdaProbabilities}
\end{figure}

According to Eq.~\ref{logProbC}, three histograms are required to calculate $\Lambda$. The $N\textsubscript{obs}$ values come from the raw data histograms. The $N\textsubscript{back}$ values come from the shadow-free averaged background histograms. Those histograms show backgrounds near the Sun, averaged over 600 skies for the whole data set and 3000 skies for the min/max data sets. In each case, the Sun is not present, and therefore no shadow is evident. There is a distinct decrease in particle density for higher $\Delta$DEC due to the detector's position at a relatively high latitude, but otherwise the distributions appear flat. The $N\textsubscript{th}$ values come from the shadow template multiplied by the $N\textsubscript{back}$ values. The most likely location of the shadow is wherever $\Lambda$ is maximized, $\Lambda\textsubscript{max}$. This corresponds to the location where the data histogram agrees best with the shadow template. The results are shown in Fig.~\ref{fig:lambdaProbabilities} along with the Moon shadow located by the same methods.

The positions of the shadows are unique from one another, and distorted in various ways. The bin sizes in the histograms are all $0.08^{\circ}$ so the only possible shadow locations are the bin centers. The full data set seems to be a combination of the solar max and solar min shifts, and the solar max/min are shifted in distinctly different directions. This may be due to a different polarity ratio in the field as is discussed in Section 3.6. The primary goal of the magnetic field simulations is to see if their differences can be reproduced by varying the Parker spiral model. To determine confidence in the shadow locations, probabilities must be simulated using Monte Carlo techniques.

\subsection{Confidence in the Shadow Locations}

\begin{figure} [htbp] 
    \centering
    \includegraphics[width=1.0\textwidth]{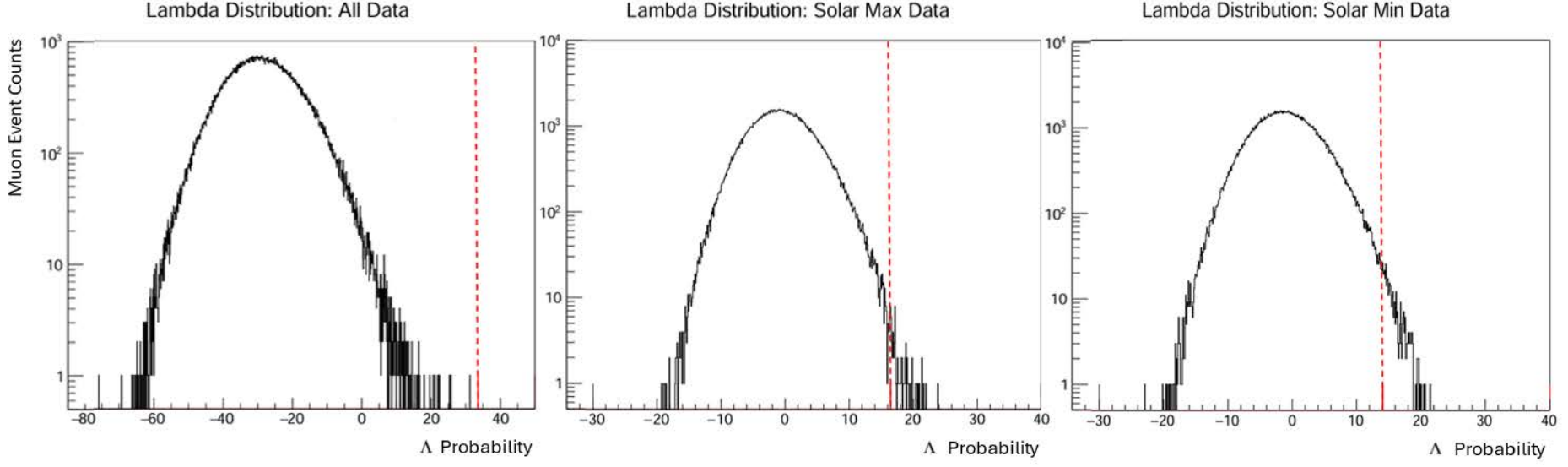}
    \caption[All Data Lambda Distribution]{$\Lambda\textsubscript{max}$ distribution over the full data range (left), the solar max range (center), and the solar min range (right) for the Sun shadow. The red lines mark the $\Lambda\textsubscript{max}$ value from the real data set. No randomly generated values were greater than the observed $\Lambda\textsubscript{max} = 33.57$ in the full data distribution, the observed $\Lambda\textsubscript{max} = 16.44$ in the solar max distribution, or the observed $\Lambda\textsubscript{max} = 14.01$ in the solar min distribution. There were no events greater than the observed $\Lambda\textsubscript{max} = 33.57$ in the full data distribution. There were, however, 67 randomly generated values were greater than the observed $\Lambda\textsubscript{max} = 16.44$ in the solar max distribution and 401 randomly generated values were greater than the observed $\Lambda\textsubscript{max} = 14.01$ in the solar min distribution. These correspond to probabilities of $3.7 \times 10^{-4}$ and $2.2 \times 10^{-3}$. The probability for occurrence of such rare events can also be estimated by using the distributions' means and standard deviations. The full data range event is $6.01\sigma$ to the right of the mean, the solar max event is $3.59\sigma$ to the right of the mean, and the solar min event is $3.06\sigma$ to the right of the mean.}
    \label{fig:LambdaCharts}
\end{figure}

$\Lambda\textsubscript{max}$ values were determined from data files being compared to background histograms. Each simulated sky is subject to Poissonian fluctuations as mentioned in Section 2.2. This means it is possible that a particular section of the sky mimics the appearance of the shadow. To determine how likely random fluctuations are to create a shadow-like $\Lambda$, off-source regions of the data are examined using the likelihood's $\chi^{2}$ distribution \cite{Cash:1979}. Simulated data sets were produced by sampling the respective background histograms for each of the three data periods. The number of muons in each data period was equal to the number of particles within $4^{\circ}$ of the Sun in the real data. Each time a histogram was generated, nine $\Lambda\textsubscript{max}$ values were located; one each in 20 by 20 bin gridded sections of the histogram. These $1.6^{\circ} \times 1.6^{\circ}$ sections are slightly smaller than the shadow template providing the $N\textsubscript{th}$ values which is approximately $2.0^{\circ}$ in diameter. This allows for multiple maximum values to be located within each data set while also permitting minimal overlap between gridded sections. A total of $2 \times 10^{4}$ random histograms were generated for each scenario which produces $1.8 \times 10^{5}$ $\Lambda\textsubscript{max}$ values per data period. These values were plotted as distributions to show what $\Lambda\textsubscript{max}$ values are expected to occur due to random fluctuations alone. The histogram for the whole dataset, max dataset, and min dataset are shown in Fig. \ref{fig:LambdaCharts}.

The fact that the observed $\Lambda\textsubscript{max}$ value for the full dataset is $\unit[6.01]{\sigma}$ away from the mean can be used to generate a probability of random occurrence. Using the error function one can estimate the probability with the following formula

\begin{equation}
\centering
 P(x) = 1 - \mathrm{erf}(\frac{x}{\sqrt{2}}),
 \label{sigmaProb}
\end{equation} where $x$ is the number of sigma away from the mean of the observed value and $\mathrm{erf}$ is the error function. This gives a result of $ P(6.01) = 1.86 \times 10^{-9}$. This probability can be further reduced by considering that half of the values outside of $6.01\sigma$ are on the opposite side of the distribution so dividing by 2 gives the result $P = 9.3 \times 10^{-10}$. 

In addition, the observed shadow occurs in the central section of the nine-section division of the $\Lambda$ probability histogram. This is its expected location. Since the distribution was produced from nine square sections per histogram, $\frac{8}{9}$ of these instances occur outside the central section. Therefore dividing by 9 reaches the final probability value of $1.0 \times 10^{-10}$. This corresponds to an expectation of $1.86 \times 10^{-5}$ events for every $1.8 \times 10^{5}$ simulated skies so it is no surprise that there were no values of that magnitude randomly generated.
 
For the solar min and max datasets, histograms were generated that contained $\Lambda\textsubscript{max}$ values greater than the observed values, as is evident in Fig. \ref{fig:LambdaCharts}. They show that the solar max $\Lambda\textsubscript{max}$ value has a fractional probability $P\textsubscript{s-max} = 3.7 \times 10^{-4}$ and the solar min value has a fractional probability $P\textsubscript{s-min} = 2.2 \times 10^{-3}$ of occurring randomly in the absence of a source. These probabilities are determined only by $\Lambda\textsubscript{max}$ values greater than the observed $\Lambda\textsubscript{max}$ so dividing by 2 is redundant. Dividing by 9 possible sections further reduces the fractional probabilities to $P\textsubscript{s-max} = 4.1 \times 10^{-5}$ and $P\textsubscript{s-min} = 2.4 \times 10^{-4}$. The same method can be employed using $\sigma$ and the error function as was used for the full data set on solar max and min. The results for all three data sets are shown in Table \ref{shadowProbabilities}. 

\begin{table}[htbp]
\centering
\begin{tabular}{|c|c|c|c|}
    \hline
    Data Period & Fractional Probability & $\sigma$-Probability & Dev. of Obs. from Mean\\
    \hline
    Full Data & $NA$ & $1.0 \times 10^{-10}$ & $6.01\sigma$ \\
    \hline
    Solar Max & $4.1 \times 10^{-5}$ & $1.8 \times 10^{-5}$ & $3.59\sigma$ \\
    \hline
    Solar Min & $2.4 \times 10^{-4}$ & $1.2 \times 10^{-4}$ & $3.06\sigma$ \\
    \hline
\end{tabular}
\caption[Shadow Probabilities]{Probability of each $\Lambda\textsubscript{max}$ value occurring randomly. Probabilities are determined two ways; by the distribution of $\Lambda\textsubscript{max}$ values and by way of standard deviation $\sigma$ and the error function. Note that there is no fractional probability for the full data set because no simulated distributions had larger $\Lambda \textsubscript{max}$ values than the observed distribution. }
\label{shadowProbabilities}
\end{table}

Given that the full dataset has a probability on the order of $10\textsuperscript{-10}$ and that the min and max datasets are part of the full dataset, it is almost certain that the shadow is present. That said, the precise position of the shadow is less certain at solar min and solar max.

\section{Magnetic Field Models and Computational Model Building} 
\label{sec:Models}

\subsection{The Parker Spiral Model}

The solar wind is a plasma that has many properties relevant to the nature of the magnetic field within the heliosphere, including the “frozen-in” condition. The wind has a high conductivity which causes magnetic flux \(\frac{d\Phi}{dt}\) within the plasma to go to zero \cite{Kallenrode:2004plasma}. The consequence is that the magnetic field is carried by the solar wind as it propagates away from the Sun. The heliosphere is defined by the field carried with the solar wind, therefore the structure and strength of the heliosphere’s magnetic field is primarily dependent on the properties of the solar wind and, by extension, the Sun itself. 

The solar wind carries plasma away from the Sun at what is assumed to be a constant velocity $u\textsubscript{sowi}$, which will be referred to as the solar wind velocity. In reality the solar wind is not constant, but rather it has fast and slow components. The fast solar wind has a density of approximately $\unit[3]{ions/cm^{3}}$ and velocities ranging from $\unit[400]{km/s}$ to $\unit[800]{km/s}$, while the slow solar wind has a density of approximately $\unit[8]{ions/cm^{3}}$ and velocities ranging from $\unit[250]{km/s}$ to $\unit[400]{km/s}$. The intersection of these two velocity ranges is used as the constant velocity: $u\textsubscript{sowi} =\unit[400]{km/s}$. The Sun also rotates with a sidereal period of $\unit[27]{days}$ \cite{Kallenrode:2004plasma}. This is represented as an angular frequency $\omega\textsubscript{\(\odot\)}$. Therefore a plasma “parcel” moves with a radial velocity $u\textsubscript{sowi}$ and has an angular velocity of $\omega\textsubscript{\(\odot\)}$. Its position can be represented with the position vector

\begin{equation}
\centering
 \vec{r} = [r(t), \phi(t)] = [u\textsubscript{sowi}t + r\textsubscript{0}, \omega\textsubscript{\(\odot\)}t + \phi\textsubscript{0}], 
\end{equation} where the origin is the center of the Sun and the position of the parcel at $t=0$ is \( (r,\phi) = (r\textsubscript{0}, \phi\textsubscript{0})\) \cite{Kallenrode:2004plasma}. By implementing the frozen-in condition of the solar wind and reconfiguring the position equation as time independent, Gauss's Law can be used to determine the magnetic field $\vec{B}$ in the equatorial plane, \( \vec{B} = (B\textsubscript{r},B\textsubscript{\(\phi\)}) \). For sufficiently large radii $r\omega\textsubscript{\(\odot\)} > u\textsubscript{\(\phi\)}$ where $u\textsubscript{\(\phi\)}$ is the $\phi$ component of the solar wind velocity, the magnetic field can be represented as

\begin{equation}
\centering
 \vec{B} = \frac{r\textsubscript{0}\textsuperscript{2}}{r\textsuperscript{2}} B\textsubscript{0} \hat{r} - \frac{ \omega\textsubscript{\(\odot\)} r } {u\textsubscript{r}} B\textsubscript{0} \hat{\phi}. 
 \label{solarMagneticField}
\end{equation}Here u\textsubscript{r} is the radial component of the solar wind velocity and $B\textsubscript{0}$ is the magnetic field magnitude in the Sun's photosphere. This field, along with the superposition of the geomagnetic field, can be used to calculate the Lorentz force on a particle at any position $\vec{r} = [r,\phi]$ near the equatorial plane.

\subsection{The Geomagnetic Field}

The interplanetary magnetic field will dominate the geomagnetic field in most regions of the solar system except for those near the Earth. The geomagnetic field can be modeled as a dipole with moment $ M\textsubscript{E} =\unit[8\times10\textsuperscript{22}]{A m\textsuperscript{2}}$ \cite{Kallenrode:2004plasma}  and can be expressed in geomagnetic coordinates as 

\begin{equation}
\centering
     \vec{B}\textsubscript{m} = \frac{\mu\textsubscript{0}M\textsubscript{E}}{4\pi \rho\textsuperscript{3}} (-2\sin\Phi \hat{\rho} + \cos\Phi \hat{\Phi} )
     \label{geoMagneticFC}.
\end{equation}
In Eq.~\ref{solarMagneticField} the Sun-centered spherical coordinate system is expressed in vector form as $\vec{r} = (r,\theta, \phi)$. The geomagnetic coordinates introduced by Eq.~\ref{geoMagneticFC} define an Earth centered spherical coordinate system where the magnetic latitude $\Phi$ is 0 at Earth's magnetic equator. This coordinate system can be expressed as $\vec{\rho}\textsubscript{m} = (\rho, \Phi, \Lambda)$ where $\rho$ is the radial, $\Phi$ is the polar, and $\Lambda$ is the azimuthal coordinate. Note that $\Lambda$ in this case refers to a magnetic longitude, not the probability factor from earlier.  For the purposes of putting the geomagnetic field into a Sun-centered coordinate system, a second set of geographic coordinates must also be defined in which the polar axis is aligned with the Earth's rotation axis. Its position vector is expressed as $\vec{\rho}\textsubscript{g} = (\rho, \varphi, \lambda)$, where $\rho$ is the radial coordinate for both geocentric coordinate systems as one is simply a rotation of the other. Both $\Lambda$ and $\lambda$ refer to longitude in the geomagnetic and geographic coordinate systems respectively. Since longitude can be measured counter-clockwise from a top down perspective, it is equivalent to the azimuthal angle under the condition that $\Lambda , \lambda = 0^{\circ}$ E is in the same position as $\phi\textsubscript{m} , \phi\textsubscript{g} = 0$. In a similar way, polar angles must be defined for each geocentric coordinate system as $\theta\textsubscript{m} = \frac{\pi}{2} - \Phi$ and $\theta\textsubscript{g} = \frac{\pi}{2} - \varphi$ since $\Phi$ and $\phi$ are latitudes rather than true polar angles.

The location of the Earth is established in Sun-centered coordinates as $\vec{r}\textsubscript{E} = (x\textsubscript{E},y\textsubscript{E},z\textsubscript{E})$. This allows for the convenient vector relation $\vec{r} = \vec{\rho}\textsubscript{m},\textsubscript{g} + \vec{r}\textsubscript{E}$ for a particle in the sun-centered coordinate system. The Earth orbits on an elliptical path with the Sun at one focal point \cite{Kallenrode:2004plasma}. Therefore, its position vector in Sun centered coordinates can be represented by the vector equation

\begin{equation}
\centering
    \vec{r}\textsubscript{E}(T) = (a \cdot \cos\frac{2\pi T}{P} + c)\hat{x} + b \cdot \sin\frac{2\pi T}{P}\hat{y}
    \label{earthPosition},
\end{equation}
where $T$ is the number of days after aphelion. Aphelion occurs on day 185.5833 (July 5, 8:00 AM GMT) on average each year. The various orbital parameters of the Sun/Earth system are shown in Table \ref{orbitalParameters} below.

\begin{table}[htbp]
\centering
\begin{tabular}{|c|c|c|c|}
    \hline
    $P$ (period) & $a$ (semi-major axis) & $b$ (semi-minor axis) & $c$ (focal distance) \\
    \hline
    365.25 days & $1.49598 \times 10\textsuperscript{11}$ m & $1.49577 \times 10\textsuperscript{11} $ m & $2.49957 \times 10\textsuperscript{9} $ m \\
    \hline
\end{tabular}
\caption[Orbital Parameter Values]{Values of the various orbital parameters of the Earth-Sun system. The Sun is located at one focal point while the Earth orbits periodically along this path.}
\label{orbitalParameters}
\end{table}

Using several geometric transformations and the location of the boreal magnetic pole as $(\varphi\textsubscript{0}, \lambda\textsubscript{0}) = (\unit[78.3]{^{\circ}N}, \unit[291]{^{\circ}E})$ \cite{Kallenrode:2004plasma}, the definition of the geomagnetic field in geographic coordinates can be represented as: 

\begin{equation}
\begin{split}
    \vec{B}\textsubscript{g} =  B\textsubscript{xm}\hat{x}\textsubscript{g} + 
    [ B\textsubscript{ym}\cos(11.7^{\circ}) + B\textsubscript{zm}\sin(11.7^{\circ}) ] \hat{y}\textsubscript{g} \:+
    \\
    [ -B\textsubscript{ym}\sin(11.7^{\circ}) + B\textsubscript{zm}\cos(11.7^{\circ}) ]\hat{z}\textsubscript{g}
    \label{geoGraphicFC}.
\end{split}
\end{equation}
The values $(B\textsubscript{xm}, B\textsubscript{ym}, B\textsubscript{zm})$ refer to the cartesian components in the geomagnetic coordinate system for the magnetic field $\vec{B}$. Several factors including the precession of the Earth and the day of the year that the summer solstice occurs leads to the definition of geographic coordinates $(x\textsubscript{g}, y\textsubscript{g}, z\textsubscript{g})$ in terms of the solar center coordinates $(x, y, z)$ as follows:

\begin{equation}
\begin{split}
    \hat{x}\textsubscript{g} = [\cos\beta \cos\alpha \sin(\frac{\pi}{2} + \delta\textsubscript{max}) + \sin\beta cos(\alpha + \frac{\pi}{2})]\hat{x}
    \\
    +\: [\cos\beta \sin\alpha \sin(\frac{\pi}{2} + \delta\textsubscript{max}) + \sin\beta \sin(\alpha + \frac{\pi}{2})]\hat{y}
    \\
    +\: \cos\beta \cos(\frac{\pi}{2} + \delta\textsubscript{max})\hat{z}
    \label{xGeographic},
\end{split}
\end{equation}

\begin{equation}
\begin{split}
    \hat{y}\textsubscript{g} = [-\sin\beta \cos\alpha \sin(\frac{\pi}{2} + \delta\textsubscript{max}) + \cos\beta \cos(\alpha + \frac{\pi}{2})]\hat{x}
    \\
    +\: [-\sin\beta \sin\alpha \sin(\frac{\pi}{2} + \delta\textsubscript{max}) + \cos\beta \sin(\alpha + \frac{\pi}{2})]\hat{y}
    \\
    -\: \sin\beta \cos(\frac{\pi}{2} + \delta\textsubscript{max})\hat{z}
    \label{yGeographic},
\end{split}
\end{equation}

\begin{equation}
    \hat{z}\textsubscript{g} = \cos\alpha \sin\delta\textsubscript{max}\hat{x} + \sin\delta \sin\delta\textsubscript{max}\hat{y} + \cos\delta\textsubscript{max}\hat{z}
    \label{zGeographic},
\end{equation}
where $\beta$ is the Earth's angle of rotation based on the time of day, $\alpha$ is the projection vector of the Earth's $z\textsubscript{g}$ axis in its orbital plane, and $\delta\textsubscript{max}$ is its maximum declination. The last set of equations necessary to determine the local geomagnetic field are the following equations to convert from geographic coordinates to geomagnetic coordinates \cite{Kallenrode:2004plasma}.

\begin{equation}
    \sin\Phi = \sin\varphi \sin\varphi\textsubscript{0} + \cos\varphi \cos\varphi\textsubscript{0} 
    \cos(\lambda - \lambda\textsubscript{0})
    \label{latConversion},
\end{equation}

\begin{equation}
    \sin\Lambda = \frac{\cos\varphi \sin(\lambda - \lambda\textsubscript{0})}{\cos\Phi}
    \label{longConversion}.
\end{equation}

The dot product in terms of cosine along with the typical arctan calculation of the azimuthal angle $(\varphi, \lambda)$ can be determined as

\begin{equation}
    \varphi = \frac{\pi}{2} - \arccos\frac{\vec{\rho} \cdot \hat{z}\textsubscript{g}}{|\rho|},
\end{equation}

\begin{equation}
    \lambda = (\lambda\textsubscript{0} - \frac{\pi}{2}) + \arctan\frac{\vec{\rho} \cdot \hat{y}\textsubscript{g}}{\vec{\rho} \cdot \hat{x}\textsubscript{g}}.
\end{equation}
With these equations, it is possible to determine the theoretical strength of the geomagnetic field in solar coordinates.

\subsection{Modeling Particle Motion}

As is the case with solid conductors, plasma also has the convenient property of internally flattening potential gradients $ [\frac{\partial V} {\partial r} = 0] $. This implies that the $\vec{E}$ field goes to 0 on length scales where its effect would be significant on a $\unit[10]{TeV}$ cosmic ray \cite{Kallenrode:2004plasma}. The fact that these GCRs are in the relativistic energy regime must also be considered. Under the condition that $\vec{E}$ = 0, the Lorentz force can be expressed as the derivative of the momentum vector $\vec{p}$ in the following manner:

\begin{equation}
\centering
 \frac{d\vec{p}}{dt} = q(\vec{v} \cdot \vec{B}). 
 \label{basicLorentz}
\end{equation}
The momentum of a relativistic particle in terms of the energy of the particle $E$ is

\begin{equation}
\centering
 \vec{p} = \left[\frac{E} {m c\textsuperscript{2}} + 1 \right]  m \vec{v} = \gamma  m \vec{v}.
 \label{momentumOfParticle}
\end{equation}
$E$ is time-dependent if work is being done on the particle. In the case of CRs moving through the HMF, the only significant force acting is the magnetic force which does no work on charged particles. This means $E$ is effectively time-independent. Substituting Eq.~\ref{momentumOfParticle} into Eq.~\ref{basicLorentz} and simplifying the result is

\begin{equation}
\centering
 \frac{d\vec{v}}{dt} = \frac{d^{2} \vec{r}}{dt} = \frac{q}{\gamma m} (\vec{v} \cdot \vec{B}).
 \label{mainDifferential}
\end{equation}
This expression will serve as the differential equation for the computational techniques discussed in Section 3.4. Note that the $\vec{B}$ in this equation is the superposition of the fields from Section 3.1 and Section 3.2.

\subsection{Simulation Strategy}

Each simulation set will be referred to as ``scenario'', and each scenario will have a corresponding ``shadow sampling template''. The shadow sampling templates provide a way to generate particles in a particular distribution and walk them backward in time using the time reversal techniques described earlier. The scenarios are divided by data periods and shadow structures. The data periods are: 1. Full Data (July 2003 - July 2016); 2. Solar Max Data (January 2012 - January 2014); 3. Solar Min Data (January 2007 - January 2009).


The shadow structure determines which type of template is used. The shadow template used earlier to determine the $\Lambda$ probability histograms shows a shadow that is circularly symmetric and partially filled in. The most likely location of the shadow is known, so a round shadow will be centered at the appropriate ($\Delta$RA, $\Delta$DEC) coordinates with respect to the Sun's position on the sky. The shadow is not necessarily a perfect circle when it reaches the rock layers. This is because the magnetic fields that deflect the GCRs could distort the round shape if certain boundaries are influenced more than others, so a distorted shadow template is also used. There is also a null hypothesis template, which is appropriate for the scenario in which the $\Lambda\textsubscript{max}$ values determined earlier are statistical fluctuations. Thus, the three shadow structures are: A. Round shadow, $0.52^{\circ}$ in diameter, centered at the $\Lambda\textsubscript{max}$ bin; B. Distorted shadow, created from the $\Lambda$ probability histogram, includes $\Lambda\textsubscript{max}$ bin; C. Flat, no shadow at all, null hypothesis.


For options A and B, the shadow will have full depth, which implies that the templates will be completely flat backgrounds with holes in them. They are holes in the sense that there would ideally be a $0\%$ chance that a particle is sampled from within that region. In the round shadow templates, this is true for any bins contained $entirely$ within the appropriate radius. Since the bins are square and the shadow border is curved, the bins on the boundary will have a probability proportional to how much of their area is enclosed by the circle. The flat backgrounds will be truly flat with equal probability.

The distorted shadow templates use the appropriate Sun $\Lambda$ probability histograms to estimate the structure of the shadows. A glance at Fig.~\ref{fig:lambdaProbabilities} reveals that the $\Lambda$ values do not decrease evenly from their maximum. A bin closer to the shadow center should incorporate a larger portion of the shadow than a distant bin. This means that the larger the $\Lambda$ value, the more likely that bin is part of the shadow. This base premise has four conditions to impose upon it for creating distorted shadow templates: 1. The $\Lambda\textsubscript{max}$ bin should have a sampling probability of 0.0, the same as the center of the rounded shadow; 2. The “volume” of a distorted shadow should be the same as that of its corresponding round shadow; 3. Only $\Lambda$ values above a certain cutoff will be included in the shadow; The shadow must be a continuous distribution. The entire shadow can be represented by a single collection of neighboring bins.

\begin{figure} [H] 
\centering
    \includegraphics[width=0.95\textwidth]{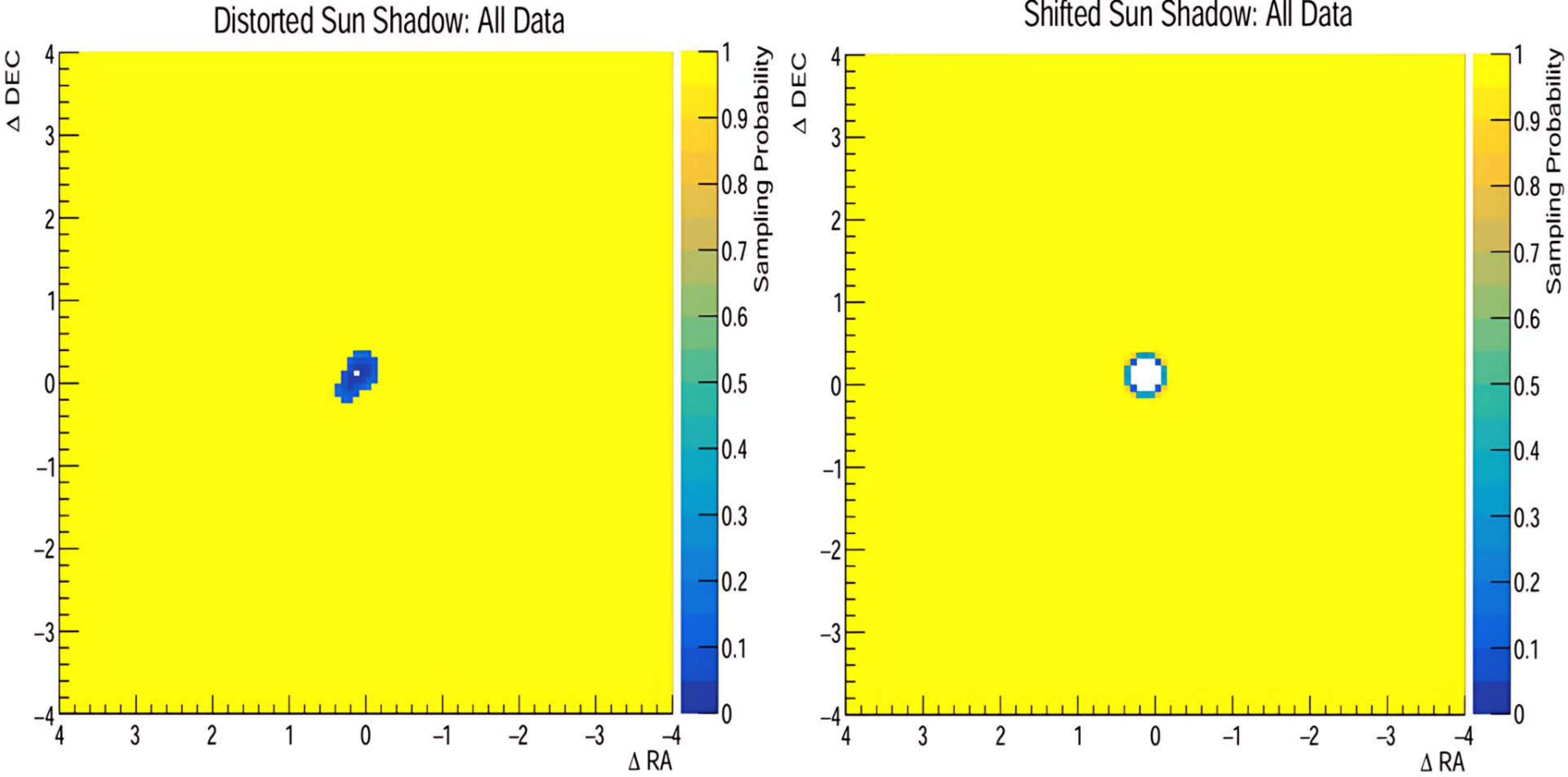}

\caption[All Data Sun Sampling]{Near Earth shadow sampling templates for the full data period (2003-2016). Each shadow has similar depth and is close to 0.0 probability across most bins. All $\Lambda$ greater than 26.79 were included to create the distorted shadow.}
\label{fig:allDataSunSampling}
\end{figure}

The volume is simply $\Sigma (1.0 - p\textsubscript{ij})$ where $p\textsubscript{ij}$ is the probability of sampling bin ij. The cutoff $\Lambda$ value is determined by progressively selecting the largest remaining bin, setting the sampling probability equal to $1 - \Lambda / \Lambda\textsubscript{max}$ , and stopping once the $(1.0 - p\textsubscript{ij})$ values sum to the area of the round shadows. It follows that only one bin will be equal to 0.0 probability (the maximum) but all of them will be relatively close to 0.0. The circular and distorted shadow sampling templates for the whole data set are shown in Fig. \ref{fig:allDataSunSampling}.

The templates for solar min and max look similar but with different shifts and distortion patterns. All three distorted shadow sampling templates required 37 bins to reach a similar shadow volume as the round shadows. The simulations will determine whether the distorted sampling templates are better models than the perfectly round templates.

\subsection{Varied Parameters}

A number of quantities relevant to the GCR motion are not constant and therefore must be varied in the time reversal simulations. They are lumped into two distinct categories: GCR parameters and solar parameters. The CR parameters have well-defined probability distributions and can be implemented by random sampling. The solar parameters are characteristically more dynamic and therefore more difficult to accurately represent.

According to Eq.~\ref{mainDifferential}, charge, mass, and kinetic energy are all relevant to GCR deflection. Prior study has shown that GCR populations have a distribution of $\approx 87\%$ protons, $\approx 12\%$ alpha particles, and $1\%$ heavier ions \cite{GCRpopulations}. This distribution provides the charge and mass of the GCRs. The $1\%$ contribution from the heavier ions has been neglected. Therefore, the CRs generated in the simulations were $87.9\%$ with charge $e$ and mass $m\textsubscript{p}$ while the other $12.1\%$ had a charge of $2e$ and mass $m\textsubscript{\(\alpha\) }$ where $e$ is the elemental charge, $m\textsubscript{p}$ is the proton mass and $m\textsubscript{\(\alpha\)}$ is the mass of an alpha particle.

Detected muons have a well-defined spectrum and can be extracted directly from the data. Prior study has shown that the MINOS far detector observes muons in a distribution with a minimum surface energy of $\unit[0.73]{TeV}$ and a peak detected value near $\unit[0.8]{TeV}$ \cite{muonEnergyDistribution}. Any observed muon corresponds to a range of primary energies characterized by a sharply peaked response curve, which makes it impossible to simply sample the energy distribution measured by the detector \cite{Gaisser:CR}. It follows that the true primary GCR energy spectrum is a continuous superposition of response curves upon a sharply peaked muon energy curve. The energy distribution of CR muons can be modeled with a power law relation \cite{Travnicek:2004PHD}. The $\unit[0.8]{TeV}$ mean of the muon curve corresponds to a mean energy of $\unit[8]{TeV}$ in the primary GCR curve \cite{muonEnergyDistribution}. In that energy regime the cosmic ray distribution follows a power law relation $E^{-2.75}$ which is enough to build a model to sample.

\begin{figure} [h] 
\centering  
    \centering
    \includegraphics[width=0.95\textwidth] {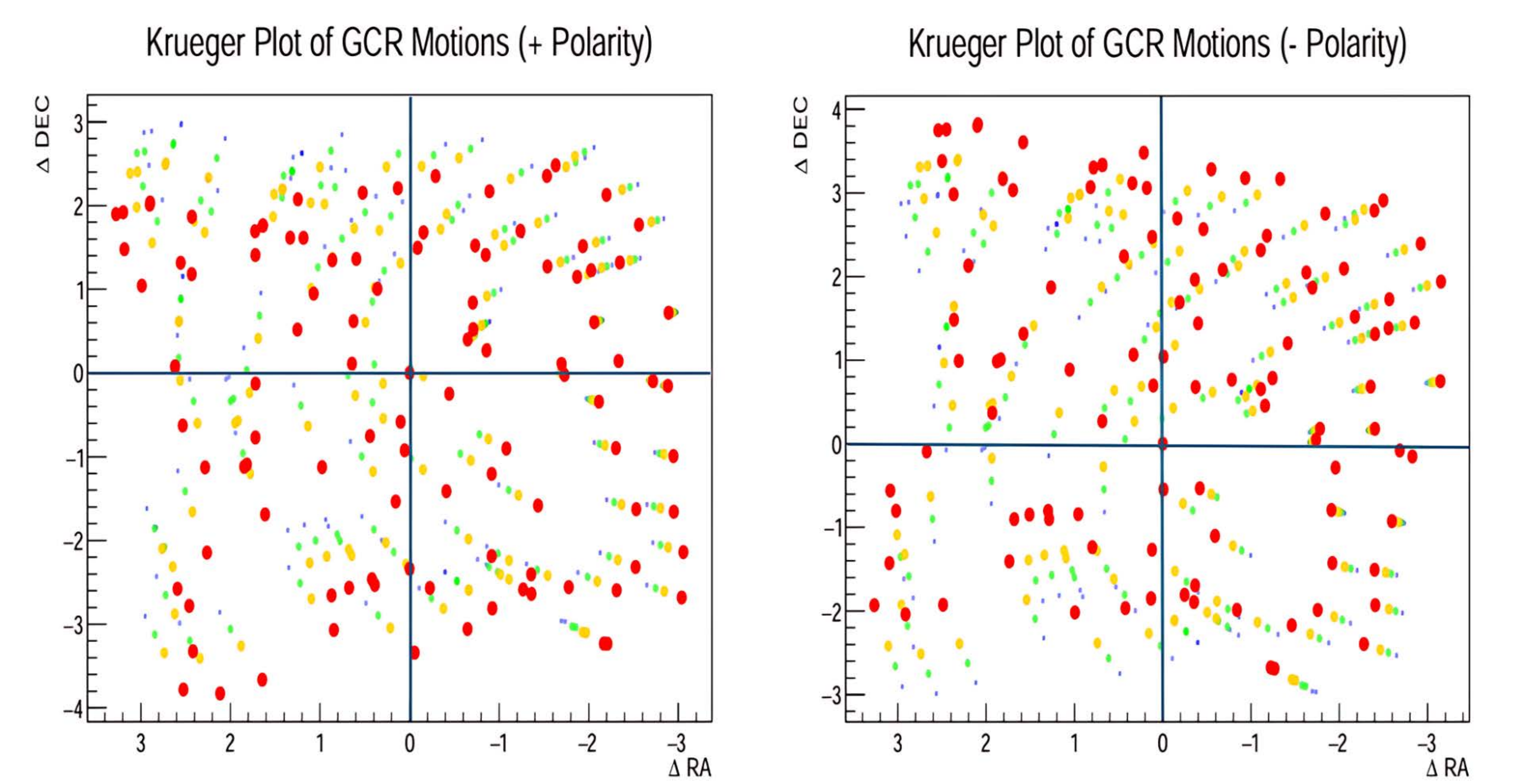}
\caption[Polarity Comparison]{Plots of GCR deflections due to positive (left) and negative (right) polarity. All GCRs in this test were assumed to have charge $e$ and mass $m\textsubscript{p}$. The smallest blue dots indicate the apparent position of particles as seen in the far detector. The larger red dots the indicate simulated original sky position when the particle was near the Sun. Intermediate green and yellow dots are transitory. Note the motions of the particles in the positive polarity field are opposite the motions in the negative polarity field. This indicates that over a period of time where neither field dominates, the effect of one field may partially mask the other.}
\label{fig:KruegerPlots}
\end{figure}

The HMF is also subject to some variation. Its various parameters have already been characterized, namely angular rotation frequency $\omega \textsubscript{\(\odot\)}$, $u\textsubscript{sowi}$, and $B\textsubscript{0}$. All of these parameters are treated as constants throughout the simulations. Their chosen values have already been motivated earlier. The HMF generally contains 2 or 4 separate polarity regions depending on magnetic field conditions in the photosphere marked by co-rotating regions at the boundaries \cite{Kallenrode:2004plasma}. Over a period of several years that the Earth spends passing through either polarity HMF is approximately equal, but over shorter time spans one polarity may dominate. For these reasons, every scenario is initially carried out with 50/50 probability for each polarity. To then compare to other polarity distributions, the ratios 33/67 and 67/33 were also used. From there, certain simulations were varied further to approach the optimum polarity ratio. An example of how each type of polarity affects GCRs in their journey between the Sun and Earth is shown in Fig.~\ref{fig:KruegerPlots}.

\section{Simulation Results and Analysis} 

\subsection{Full Field Analysis}

A 4\textsuperscript{th} order Runge-Kutta scheme was used to simulate the equation of motion over time. The step size for all simulations was -$\unit[0.1]{s}$ and the average travel time per particle in physical time was $\approx \unit[500]{s}$. Particle motions are simulated until they reach the plane that bisects the Sun and is perpendicular to the Earth's position vector. They are plotted in the coordinates ($\Delta$RA, $\Delta$DEC) from the perspective of an observer on Earth. Any particles that traveled for \unit[600]{s} were considered ``rogue'', which means they were deflected away from the plane of interception and counted separately from those plotted. Any particles that entered the Sun's photospheric radius of $\unit[6.96 \times 10^{8}]{m}$ were considered absorbed and also removed and counted separately. Since the simulation is time-reversed, a particle striking the Sun would indicate that it originated from it.

To understand what the GCR distribution looks like near the Sun a two-stage simulation was implemented. The goal of the simulation is to map any and all particles that cross the plane of intersection and then proceed toward Earth. In the first stage, particles were generated on a $\unit[28]{R\textsubscript{\(\odot\)}}$ sphere with the Sun at its center. The particle density was uniformly randomized on the entire surface of the sphere because incident GCR trajectories to the solar system are a random and uniform distribution \cite{Gaisser:CR}. The particles were given velocity vectors within $90^{\circ}$ of being directed toward the Sun. Once the particles were initialized they were stepped backward in time until they either left the $\unit[28]{R\textsubscript{\(\odot\)}}$ sphere or crossed a plane perpendicular to their initial position and bisecting Earth. The particles were then mapped using two projections, a Mollweide projection (with respect to distance) and a Mercator projection (with respect to angular distance). The result of $\unit[1.8 \times 10^{8}]{particles}$ is shown in Fig.~\ref{fig:planeOfSun} below. This result is also the motivation for the earlier assertion that initial shadow should be a nearly circular disk.

\begin{figure} [htbp] 
\centering
        \centering
        \includegraphics[width=0.95\textwidth]{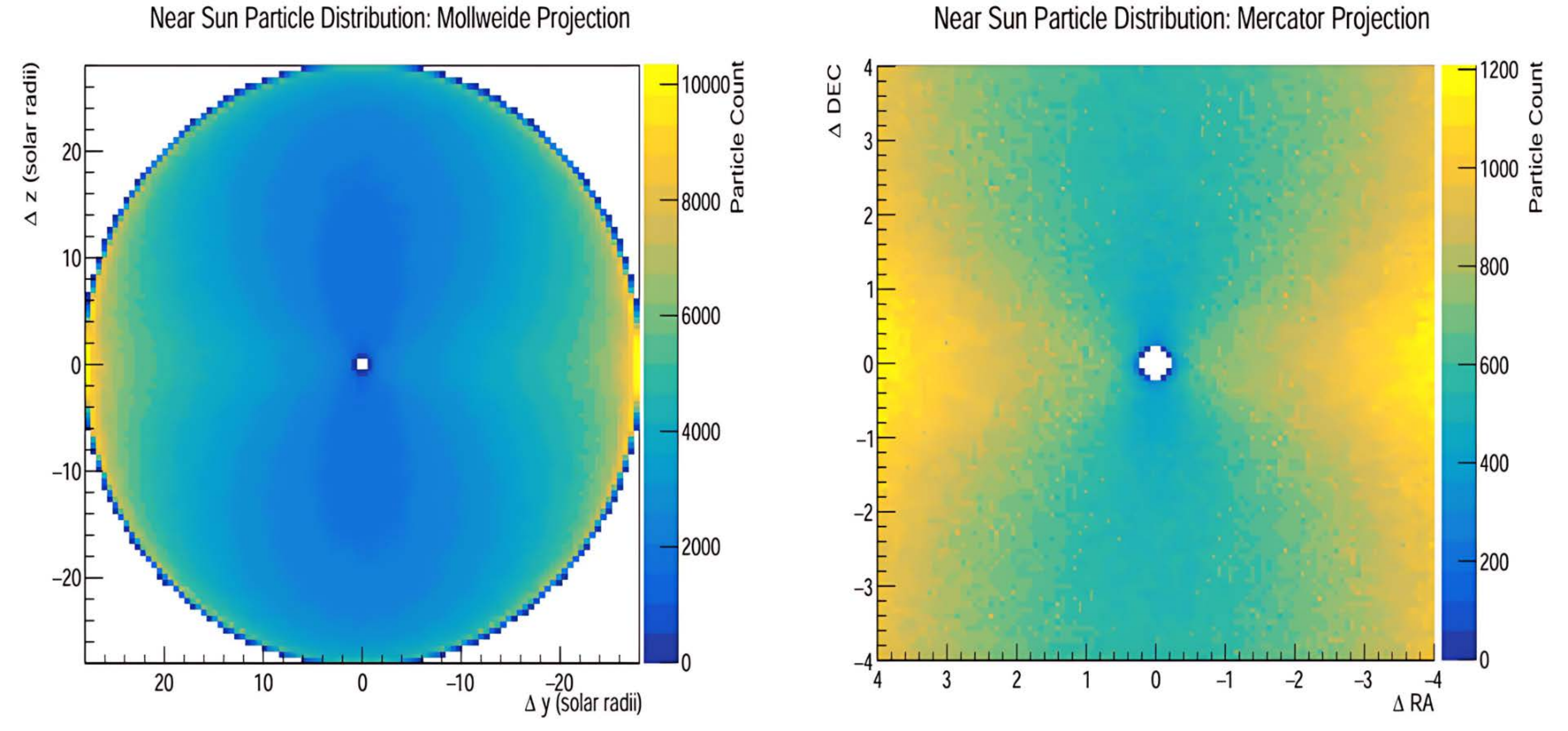} 
\caption[Plane of Sun Particle Distribution]{Distribution of particles near the Sun as they cross the plane of the Sun from the perspective of Earth. Left plot is true distance in units of solar radius. Right plot is angular distance in terms of right ascension ($\Delta$RA) and declination ($\Delta$DEC). All particles with trajectories within $90^{\circ}$ of Earth's position are included.}
\label{fig:planeOfSun}
\end{figure}

\begin{figure} [h] 
\centering
        \centering
        \includegraphics[width=0.95\textwidth]{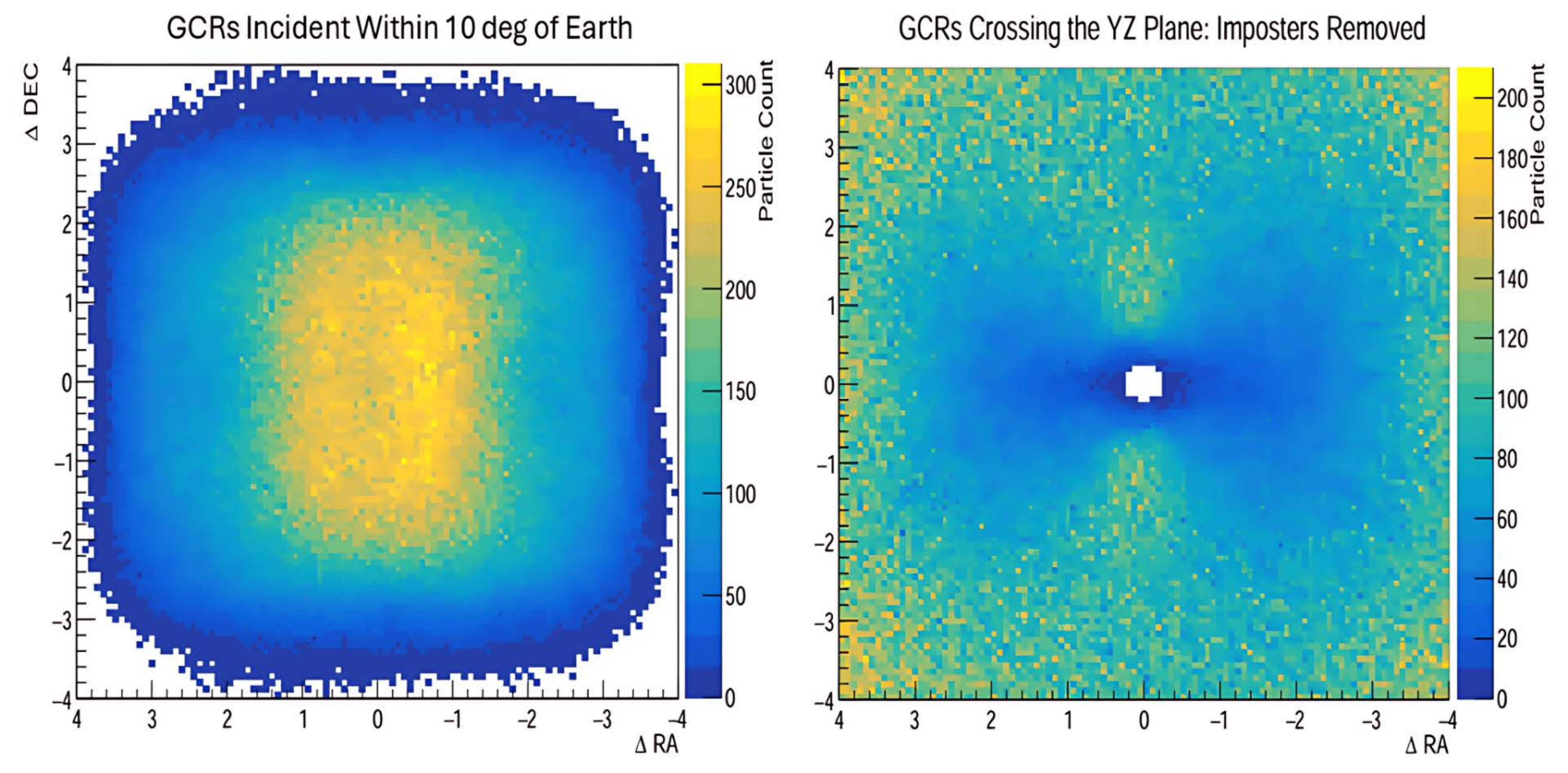}
        \label{fig:imposters}    
\caption[Imposters Removed]{Figures show the result of sampling the distribution in Fig.~\ref{fig:planeOfSun} and walking them forward 150 s in time. Particles are simulated long enough to leave the strongest parts of the HMF. The left figure shows the sampled start point of all particles with a trajectory within $10^{\circ}$ of Earth. The right figure shows where those same particles were when they crossed the plane that bisects the Sun.}
\label{fig:impostersRemoved}
\end{figure}

Imposter particles are particles that are present in the distribution but will not make it to Earth. If it was reasonable to assume that particles across this distribution had equal odds of being directed toward Earth, then this distribution could be used. The issue is that the field varies drastically in the region of space surrounding the Sun, as do the angles of the particles as they crossed the plane of interception. To remove imposters from this simulation a second stage was devised. In the first stage, the particles velocity vector relative to Earth's position was determined and stored in a 3D histogram along with the ($\Delta$RA, $\Delta$DEC) coordinates. This allows the histogram to be sampled with considerations toward the typical trajectories of the particles in particular regions of space. In the second stage, $5 \times 10^{7}$ particles were sampled and walked \unit[150]{s} forward in time. This gives them enough time to escape the stronger regions of the HMF and settle into a relatively stable trajectory as they move away from the center of the solar system. At the end of their journey, their velocity vector is again measured with respect to Earth. Only particles with trajectories within $10^{\circ}$ of Earth's position vector are included. The particles that survive this filtering process then have their $original$ angular position recorded, to show where Earth bound particles should originate from. Their angular position at the end of the simulation was also recorded as a sanity check. The results are in Fig.~\ref{fig:impostersRemoved}.

As is clear from this distribution, particles near the orbital plane are much less likely to end up directed toward Earth. The sanity check in Fig.~\ref{fig:impostersRemoved} shows the expected distribution. Particles are less dense at the edges because there are no particles outside the histogram to fill in at the edges as the distribution contracts. Otherwise, it appears to be the expected flat distribution but without the telltale signs of being measured by a high latitude underground detector.

\subsection{Residual Shadow Analysis}

\begin{figure} [h] 
\centering
        \centering
        \includegraphics[width=0.95\textwidth]{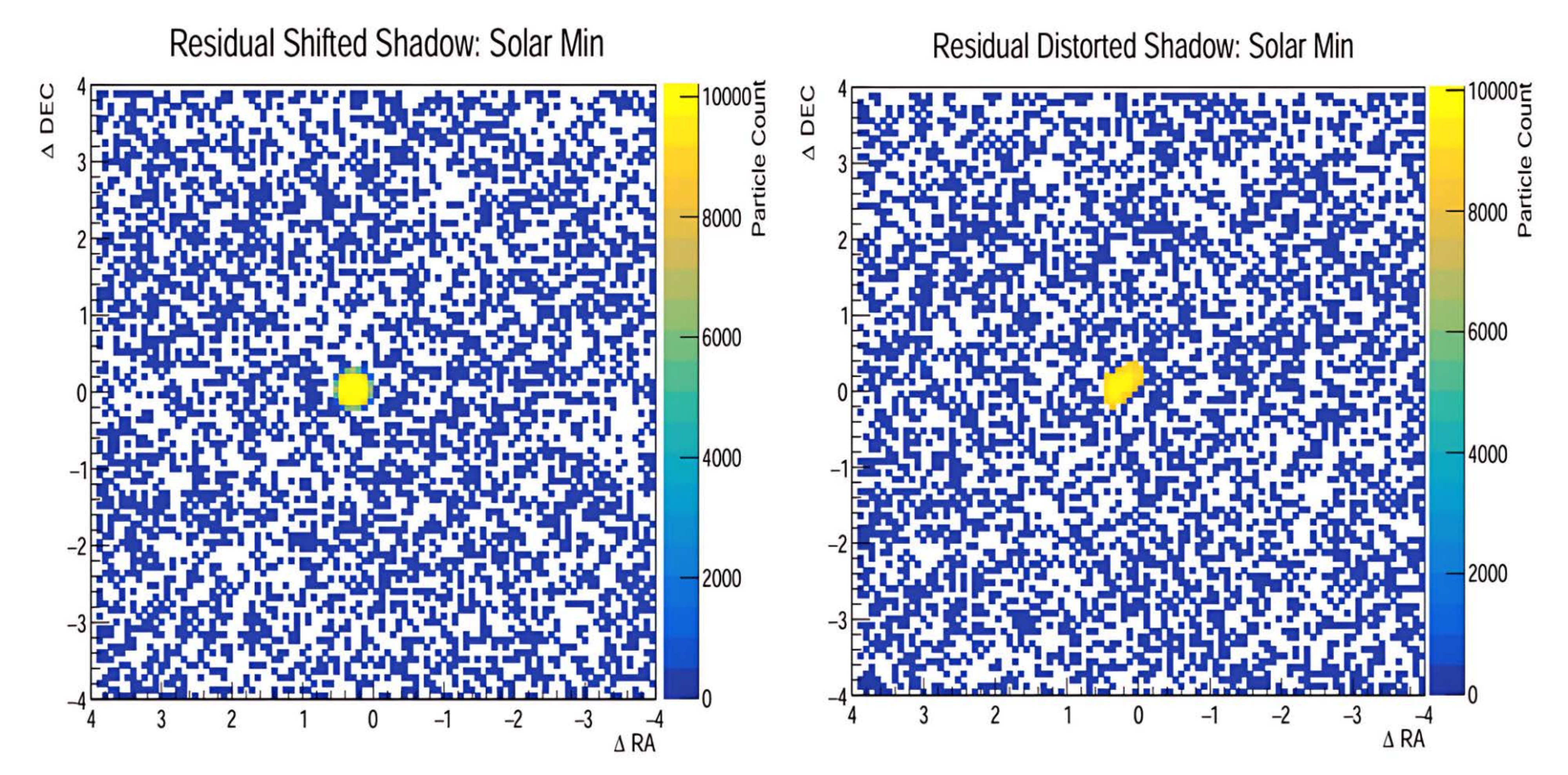}
\caption[Solar Min Residuals]{Residual sampling template produced from the sampling templates and the flat background template. The process highlights the shadow particles that are missing from the flat distribution. Each can be sampled to determine whether a particular field model could have produced them.}
\label{fig:solarMinResShadows}
\end{figure}

Residual shadow templates are created in two steps. First, apparent trajectory histograms are generated by sampling $10^{8}$ particles from the shadow templates like the one in Fig. \ref{fig:allDataSunSampling}, along with a flat template. The flat distribution then had each of the 6 shadow templates subtracted from it to create the residual shadows. Negative values were all small relative to the shadow feature and removed so the resulting histogram could be sampled. The results of this process for the solar min shadow are shown in Fig.~\ref{fig:solarMinResShadows}.

These templates allow for the shadow to be sampled directly. Recall that the shadow is a section of an otherwise flat distribution that is devoid of particles. The goal of simulating these samples is to see what the shadow particles would do if they existed. Since they were removed by striking the Sun, their ``origin'' point is the Sun itself. Any particles that do not strike the Sun should result in a distribution centered around it. A clear benefit of this type of test is that each scenario can be tested much more quickly ($\approx \unit[15]{min}$) than the full particle distributions. Only $10^{4}$ particles are necessary to get a clear idea of each distribution.

\begin{figure} [h] 
\centering
        \centering
        \includegraphics[width=0.45\textwidth]{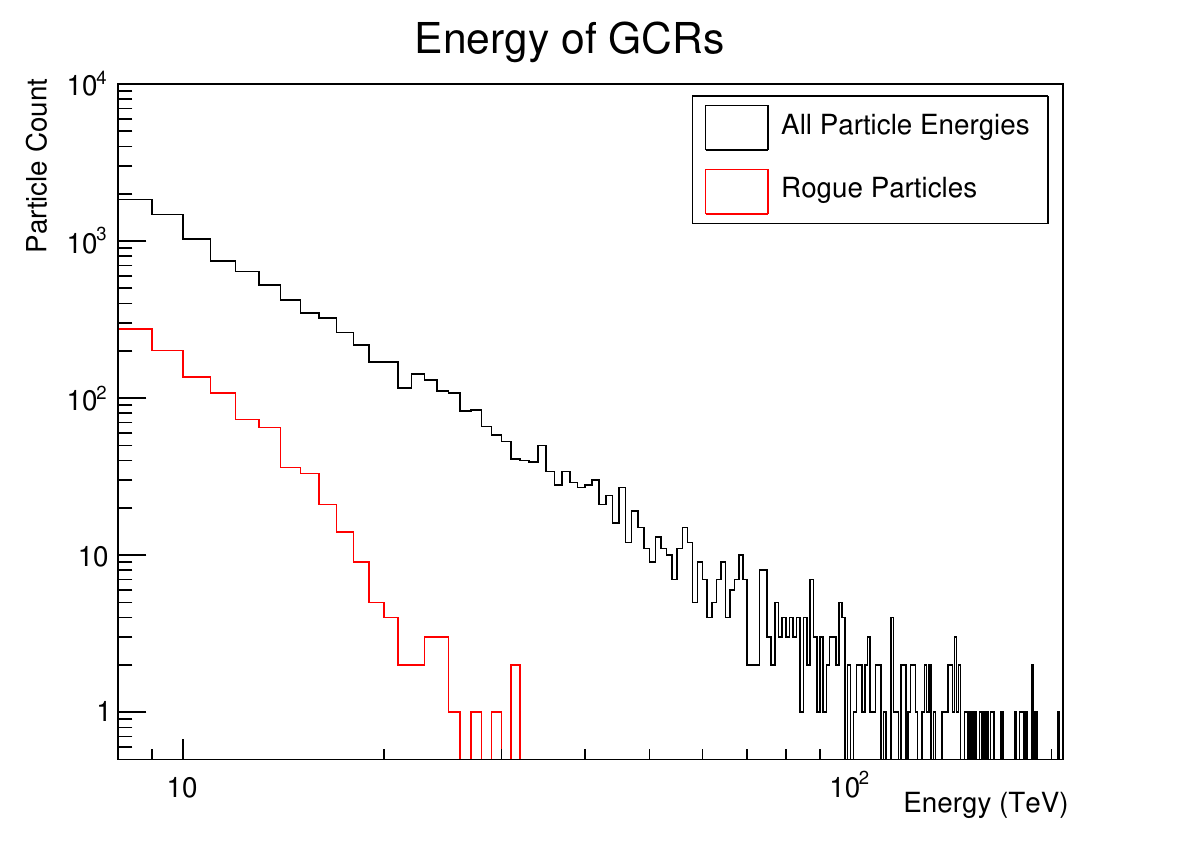}
        \label{fig:energySpectrums}
\qquad
        \centering
        \includegraphics[width=0.45\textwidth]{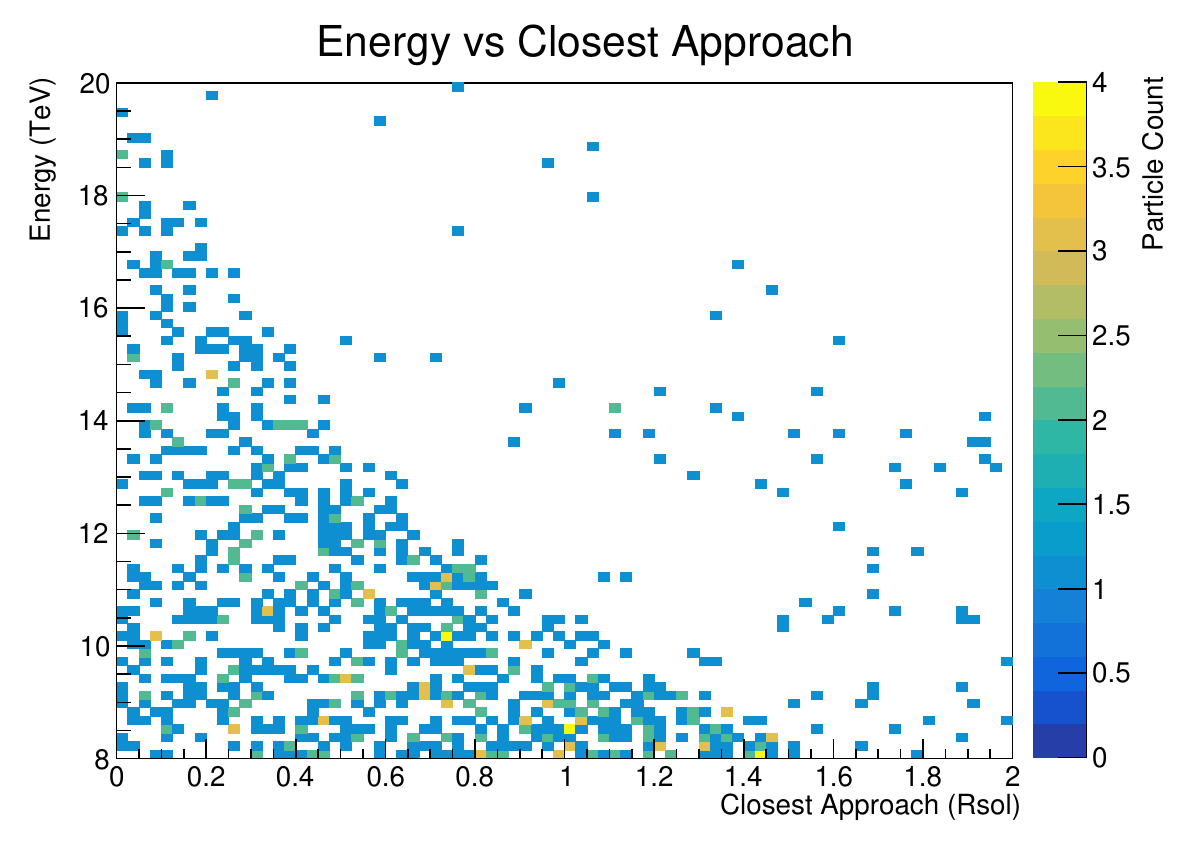}
        \label{fig:subEnergyApproach}
\caption[Energy of Rogue Particles]{Left histogram shows the energy spectrum of all sampled particles compared to the energy spectrum of particles that never crossed the plane of interception (rogue particles). It is clear that these particles are more likely to be low energy, due to how easy they are to deflect. The right plot shows the closest approach of these particles to the Sun's photosphere. Higher energy particles must come extremely close to the photosphere to be deflected.}
\label{fig:energyApproach}
\end{figure}

A brief test was undertaken to motivate the use of these ``rogue'' GCRs. If GCRs are low enough energy, they are easier to deflect away and therefore would have difficulty reaching the photosphere. The full data set with the rounded shadow structure and a 50:50 polarity ratio was tested with the typical $10^{4}$ GCRs. Every particle's closest distance from the Sun was tracked, and if it were deflected and became rogue its energy and closest approach were binned. The results of this test are in Fig.~\ref{fig:energyApproach}. They confirm that low energy particles are more likely to be completely deflected and become rogue. Also, note how quickly the energy spectrum of rogue particles falls off when compared to the overall spectrum. The main conclusion from this survey is that the number of particles actually hitting the Sun is highly dependent on the energy spectrum. 

\subsection{Initial Findings}

\begin{figure} [h] 
\centering
    \centering
    \includegraphics[width=0.95\textwidth]{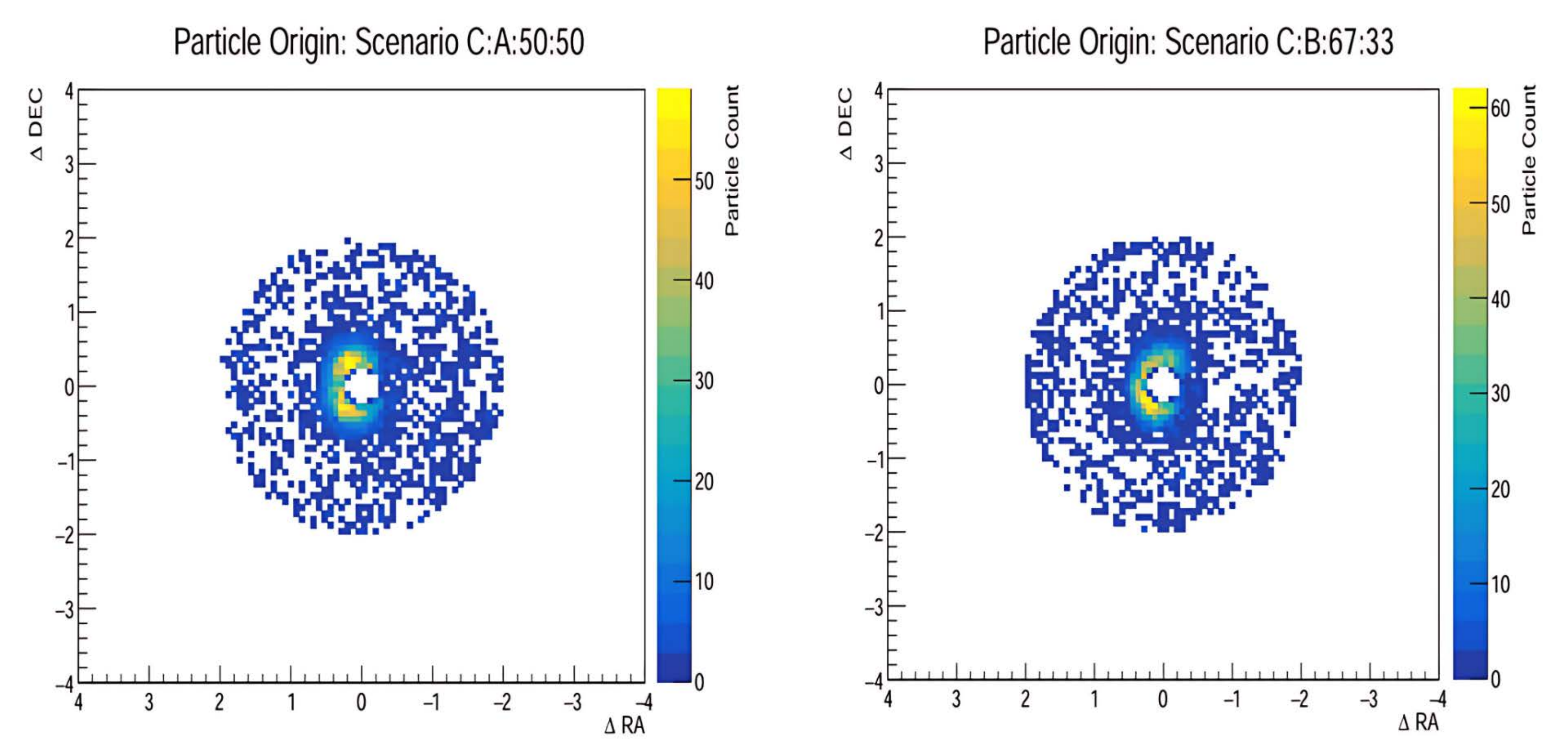}
    \label{fig:bestMinDataA}
\caption[Best Min Results]{Initial simulations in best agreement with theory from the solar minimum data set. Particles beyond $2^{\circ}$ from Sun center are removed to isolate the region of interest.}
\label{fig:bestMinData}
\end{figure}

There were 18 scenarios tested and their relevant statistics were tabulated. Shadows were either round or distorted, but they are always shifted from center. The magnetic field polarity ratios for the initial tests are 50:50, 33:67, and 67:33. This means that each of the 3 data sets had 6 potential candidate scenarios. The simulated origin distributions were considered to agree with theory based on how close their means were to (0,0). The mean was determined as a combination of the near-Sun particles and a generalized distribution of the particles that hit the Sun. Since particles that hit the Sun cannot be mapped (as they never cross the plane of intersection), it is assumed that the mean of this sub-distribution is (0.0,0.0) [i.e. the Sun's center]. The standard deviation on this distribution is assumed to be $\sigma = \frac{R\textsubscript{\(\odot\)}}{3} $ as effectively all particles are contained within both $3\sigma$ of the mean and within the Sun's disc. Overall, the solar min shadow scenarios were in best agreement with theory. The two best of these are shown in Fig.~\ref{fig:bestMinData}.

\begin{figure} [h] 
\centering
        \centering
        \includegraphics[width=0.45\textwidth]{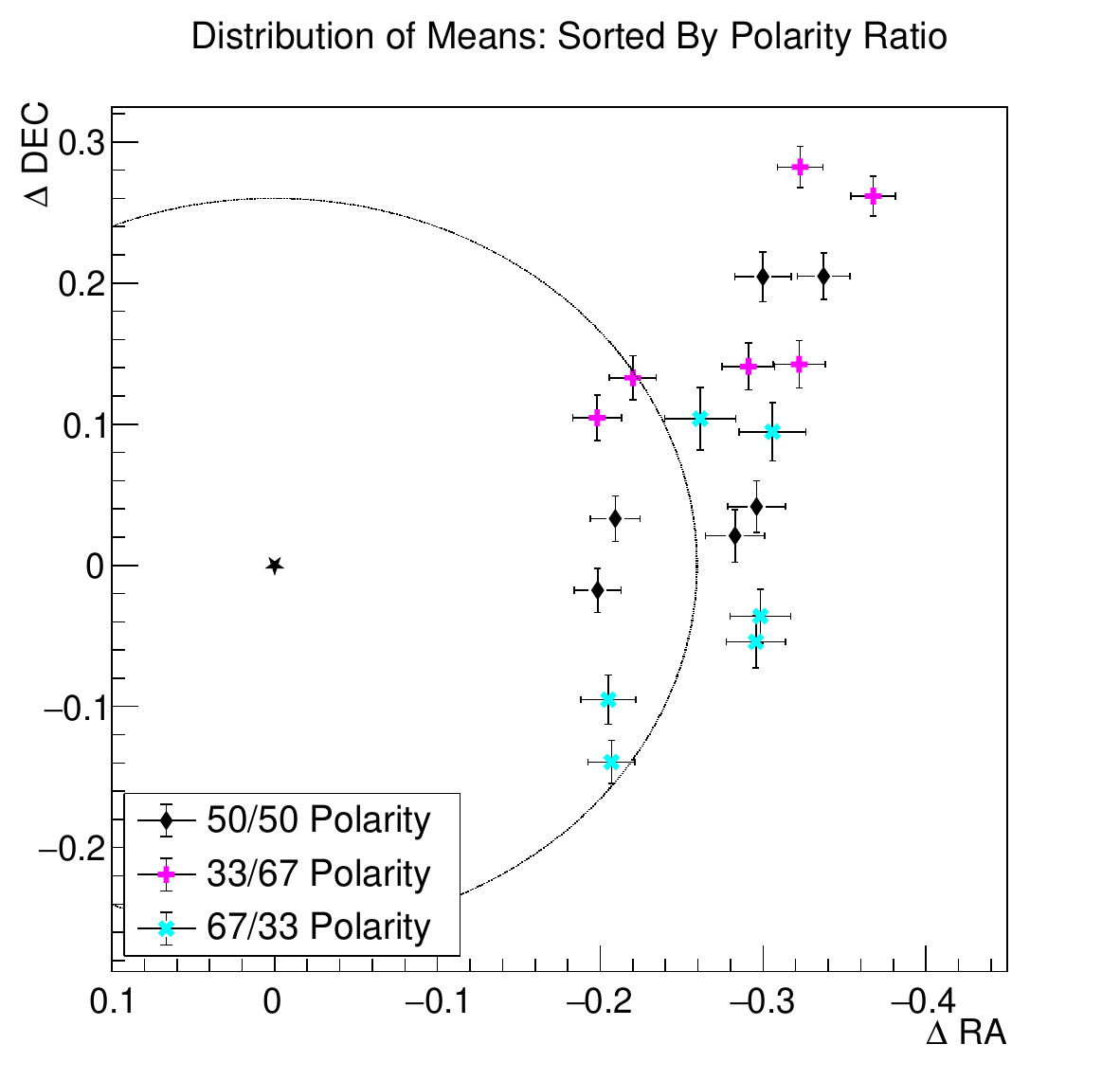}
        \label{fig:polarityMeans}
\qquad
        \centering
        \includegraphics[width=0.45\textwidth]{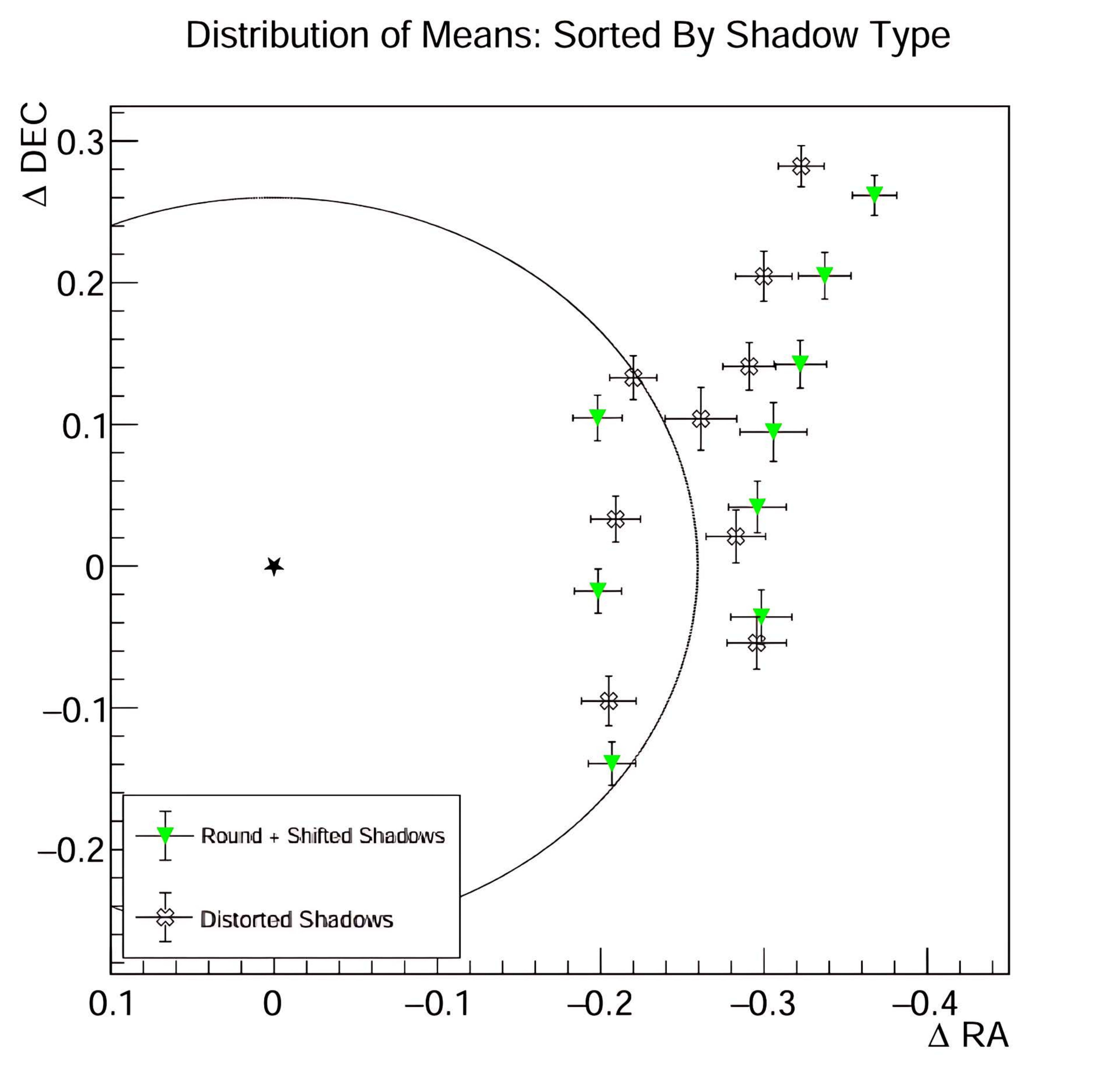}
        \label{fig:shadowTypeMeans}
\qquad
        \centering
        \hspace{-0.8cm}
        \includegraphics[width=0.45\textwidth]{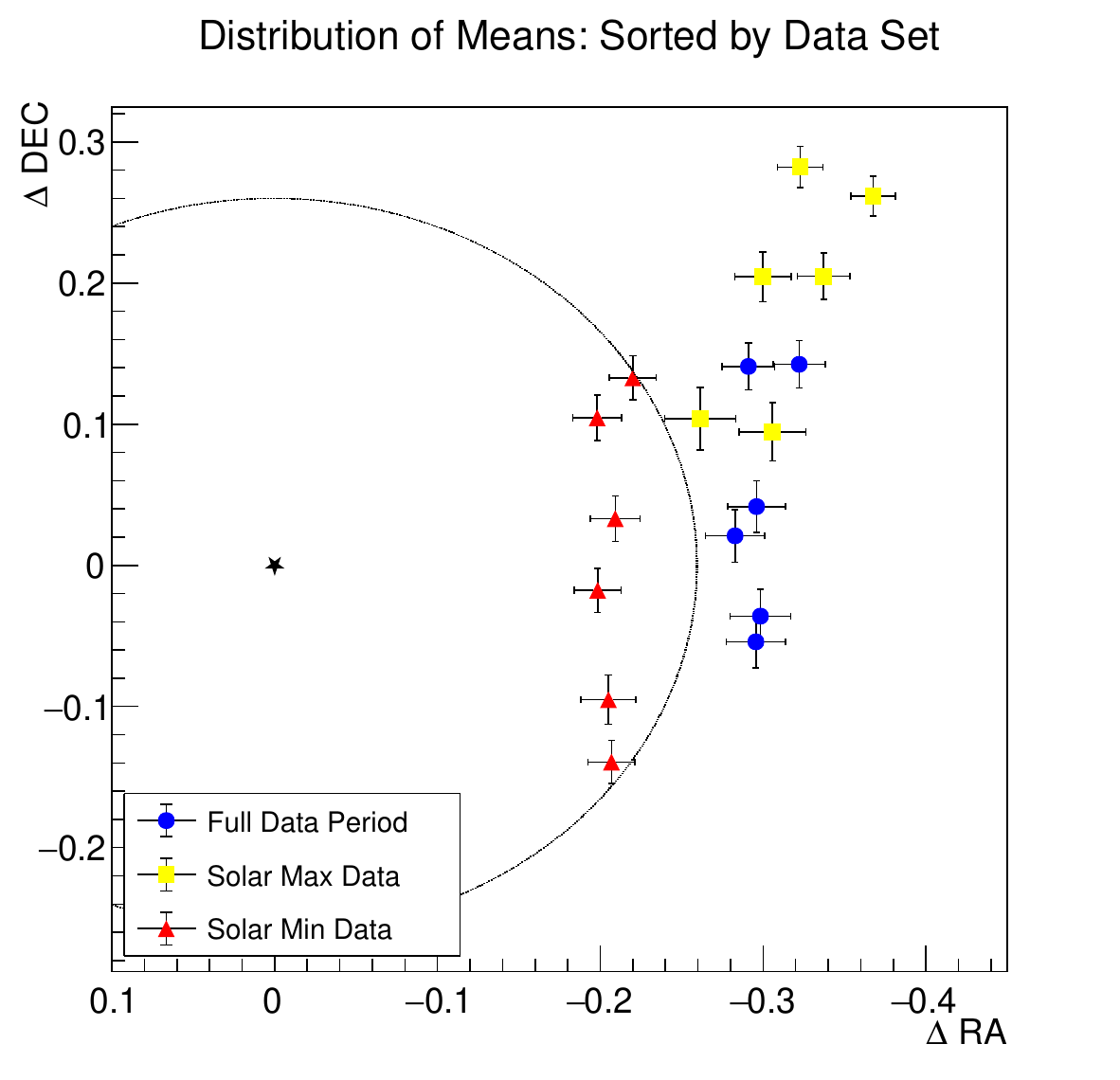}
        \label{fig:dataMeans}
\caption[Scatter Plots of Initial Data]{Scatter plot of the simulated shadow origin's $\Delta$RA and $\Delta$DEC means. The same data is plotted in all 3 figures. The top left figure categorizes the tests based on the polarity ratios. The top right figure categorizes them by which shadow template was used. The bottom figure categorizes them by the data sets. Black arc in each plot marks the the boundary of the photosphere as viewed from Earth.}
\label{fig:scatterA}
\end{figure}

All of the shadow mean centers were plotted with error bars to see how well they agree with the (0,0) mean model. The resulting plots are shown in Fig.~\ref{fig:scatterA}. They were grouped by 3 different categories. In the left figure the data were grouped by the magnetic field polarity. In the center figure the data were grouped by shadow type. Finally, in right figure the same is true except that they vary by data period.

Note the effect the polarity ratio has on the final $\Delta$DEC position of the shadow template. This behavior can be used to inform other future polarity ratios to test. It is also interesting to note that the distorted shadow templates performed better in these trials than the simple round and shifted shadows for the solar max data and the full data period, but the opposite is true for the solar min data. It seems that the success or failure of one template over the other is highly dependent on the $\Delta$RA shift. All data sets experience a right ascension shift that over-compensates for the shadow's location.

\subsection{Extended Analysis}

The $\Delta$RA and $\Delta$DEC shift for every particle in each simulation was recorded and the results show that the mean $\Delta$RA value across all scenarios is $ \mu_{\alpha} = -0.3264^{\circ} \pm 0.0152^{\circ}$. This is concerning because it indicates that all of the shadows are too close to the Sun in RA for any version of the current model. Based on prior study, the RA shift should be dominated by the geomagnetic field. Since the field is directed Northward in the Northern hemisphere, particles will tend to deflect Eastward which constitutes a positive RA shift. Recall the geomagnetic field model in Eq.~\ref{geoMagneticFC}. Using the values referenced in this paper and the coordinate conversions from Eq.~\ref{latConversion}, that model predicts that the Earth's field near the far detector (47° 49’ 13.3” N, 92° 14’ 28.5” W) has a magnitude of $|\vec{B}| = \unit[1.424\times 10^{16}] {\rho^{-3}}$ T depending on the distance from the Earth's center $\rho$. At Earth's surface this gives a $|\vec{B}|$ of approximately $\unit[5.5\times10^{4}]{nT}$. This is consistent with open source USGS data, so it seems unlikely that this is the issue. 

The energy of the simulated muons is also a significant factor in the simulation. It was motivated earlier that the GCR energy distribution is a continuous set of distributions superimposed on the cosmic ray energy distribution. They should follow the power law $\propto E^{-2.75}$ but the data may be consistent with a cutoff energy $E_{0}$ which is larger than $\unit[8.0]{TeV}$. The observed muons are known to have a cutoff surface energy of $\unit[0.73]{TeV}$ in the far detector yet the mean muon energy is $\unit[1.15]{TeV}$ \cite{muonEnergyDistribution}. Furthermore, a $\unit[1.0]{TeV}$ particle corresponds roughly to a $\unit[10]{TeV}$ cosmic ray. A larger primary GCR cutoff energy than $\unit[8.0]{TeV}$ would cause the overall distribution to deflect less overall.

\begin{figure} [h] 
\centering
        \centering
        \includegraphics[width=0.80\textwidth]{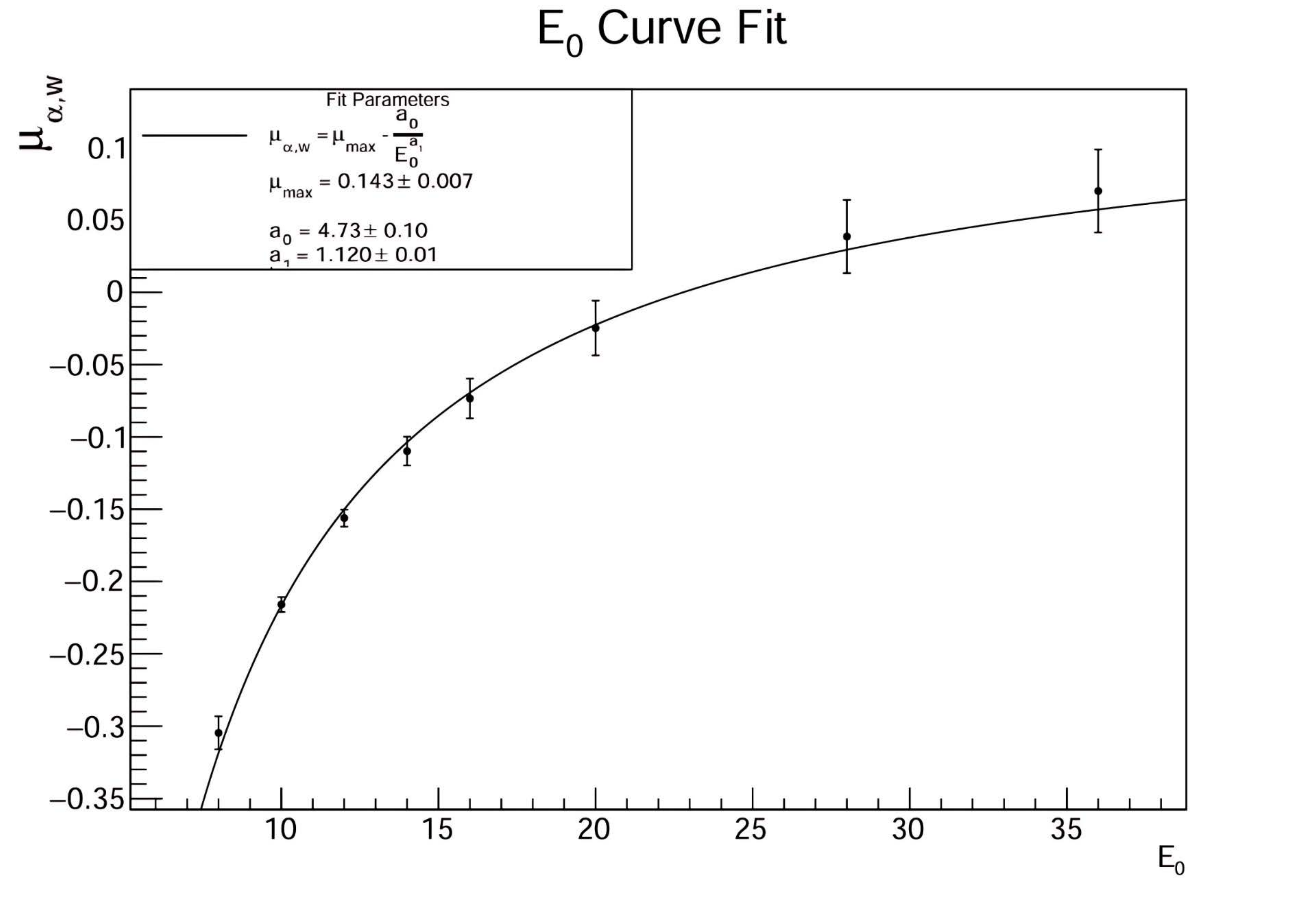}
        \label{fig:curveFit}
\qquad
        \centering
        \hspace{-0.95cm}
        \includegraphics[width=0.80\textwidth]{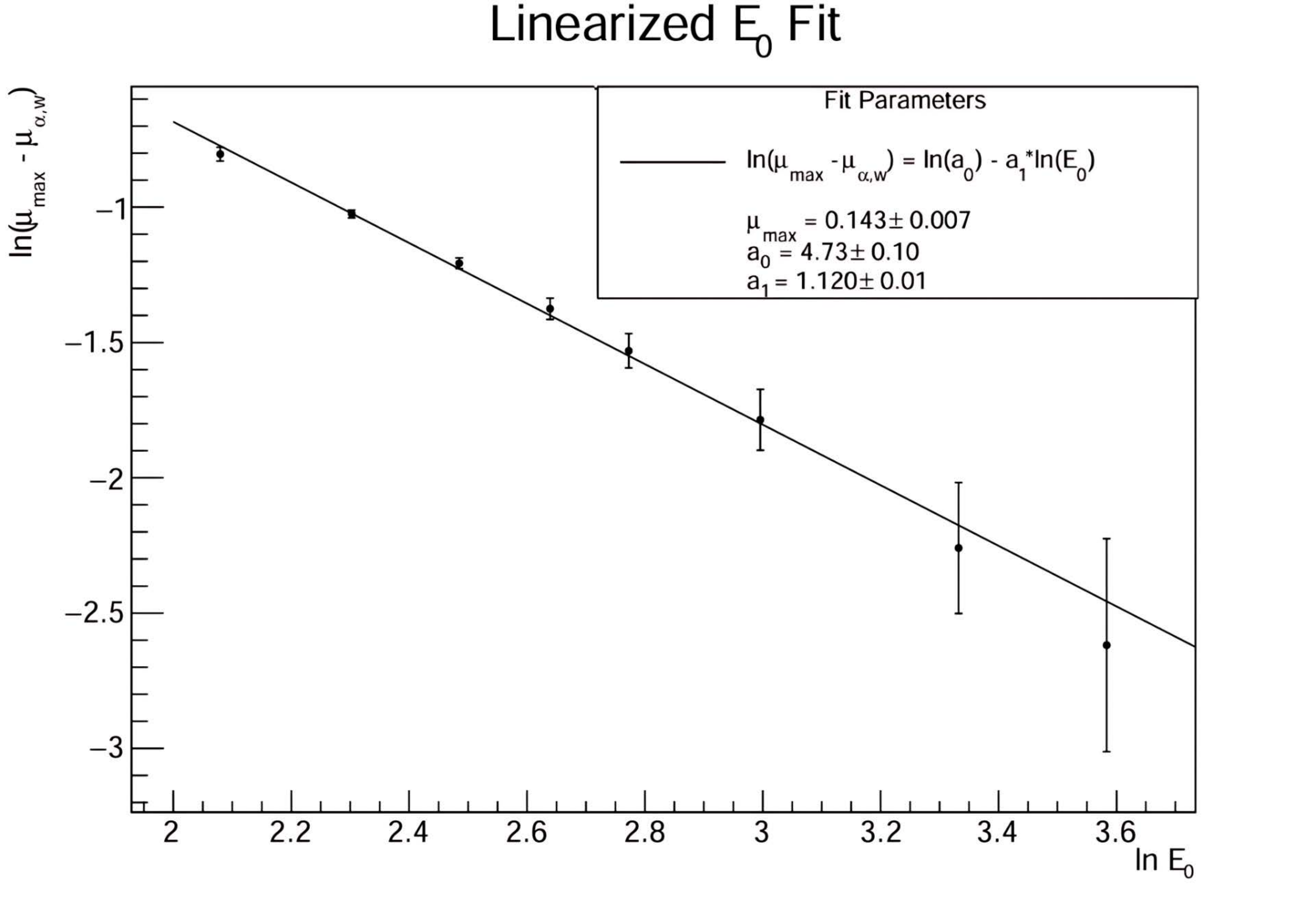}
        \label{fig:linearFit}
\caption[Energy Fit]{Fit function of the $E_{0}$ and $\mu_{\alpha,w}$ relationship. Functional form approximated using the limits of $\mu_{\alpha,w}$. The result is used to determine the cutoff energy, $E_{0}$, for which the weighted right ascension mean, $\mu_{\alpha,w}$, is 0.} 
\label{fig:energyFit}
\end{figure}

Further simulations were carried out treating the cutoff energy, $E_{0}$, as a free parameter to determine what cutoff energy is consistent with the Parker spiral model. The same 18 shadow models that were tested at \unit[8.0]{TeV} were also tested at 7 other successively increasing $E_{0}$ values. In all cases, the distribution took the form of the $E^{-2.75}$ power law with the same distribution of masses and energy and the mean right ascension $\mu_{\alpha}$ was determined. The solar min data cluster has a distinctly separate range of $\Delta$RA values from the full data and solar max periods. To prevent the latter values from outweighing the former, a weight of 2 was added to the solar min mean centers. This is justified by the fact that a single $E_{0}$ value should describe all distributions since it should not vary by month or year. A modified weighted right ascension mean $\mu_{\alpha,w}$ was calculated for each $E_{0}$ set. These values were then plotted and fit to a  function to determine where $\mu_{\alpha,w} = 0$. There is no analytical function provided from theory, but the data nevertheless follows an easily modeled form by first taking the limit of $\mu_{\alpha,w}$ as $E_{0}$ goes to 0 and $\infty$. As $E_{0}$ goes to 0, the expected GRC distributions should deflect towards $\mu_{\alpha,w} = -\infty$ at an increasing rate. As $E_{0}$ goes to $\infty$, $\mu_{\alpha,w}$ should approach some $\mu_{\alpha,max}$ because the deflection is always toward negative $\Delta$RA values and the particles would not deflect at all if they had infinite energy. By simulating all 6 data set combinations with no field acting (equivalent to infinite energy) and finding their weighted mean right ascension it was determined that $\mu_{\alpha,max} = 0.143^{\circ} \pm 0.007$. From these limits  the following functional form was constructed:

\begin{equation}
\centering
 \mu_{\alpha,w} = \mu_{\alpha,max} - a_{0}E_{0}^{-a_{1}}.
 \label{energyFit}
\end{equation}
The values $a_{0}$ and $a_{1}$ are free parameters. This function was then linearized and fit to the observed data using a least squares fit. The result of this fit is shown in Eq.~\ref{fig:energyFit}.

\begin{figure} [h] 
\centering
        \centering
        \includegraphics[width=0.45\textwidth]{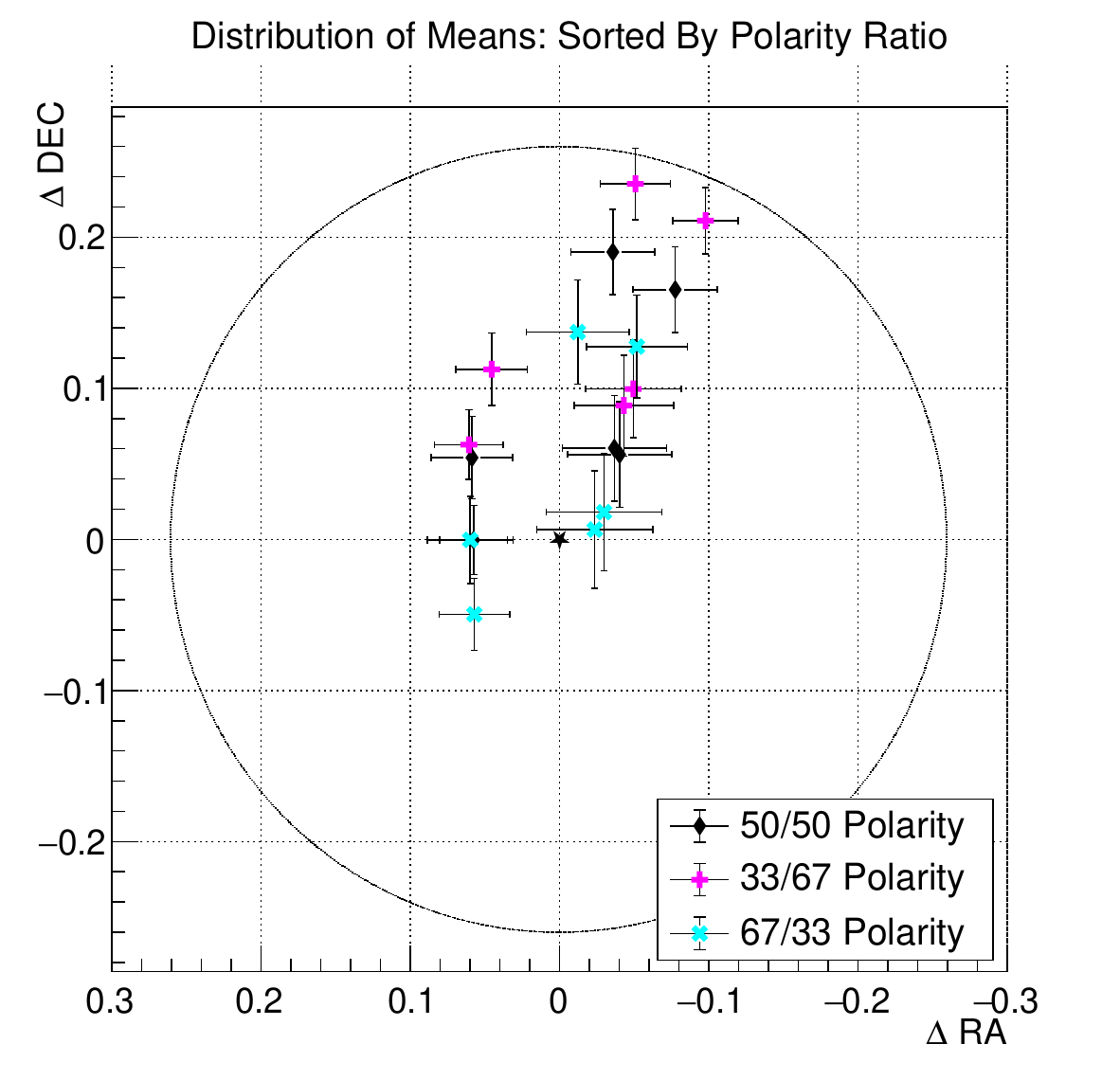}
        \label{fig:polarityMeans2}
\qquad
        \centering
        \includegraphics[width=0.45\textwidth]{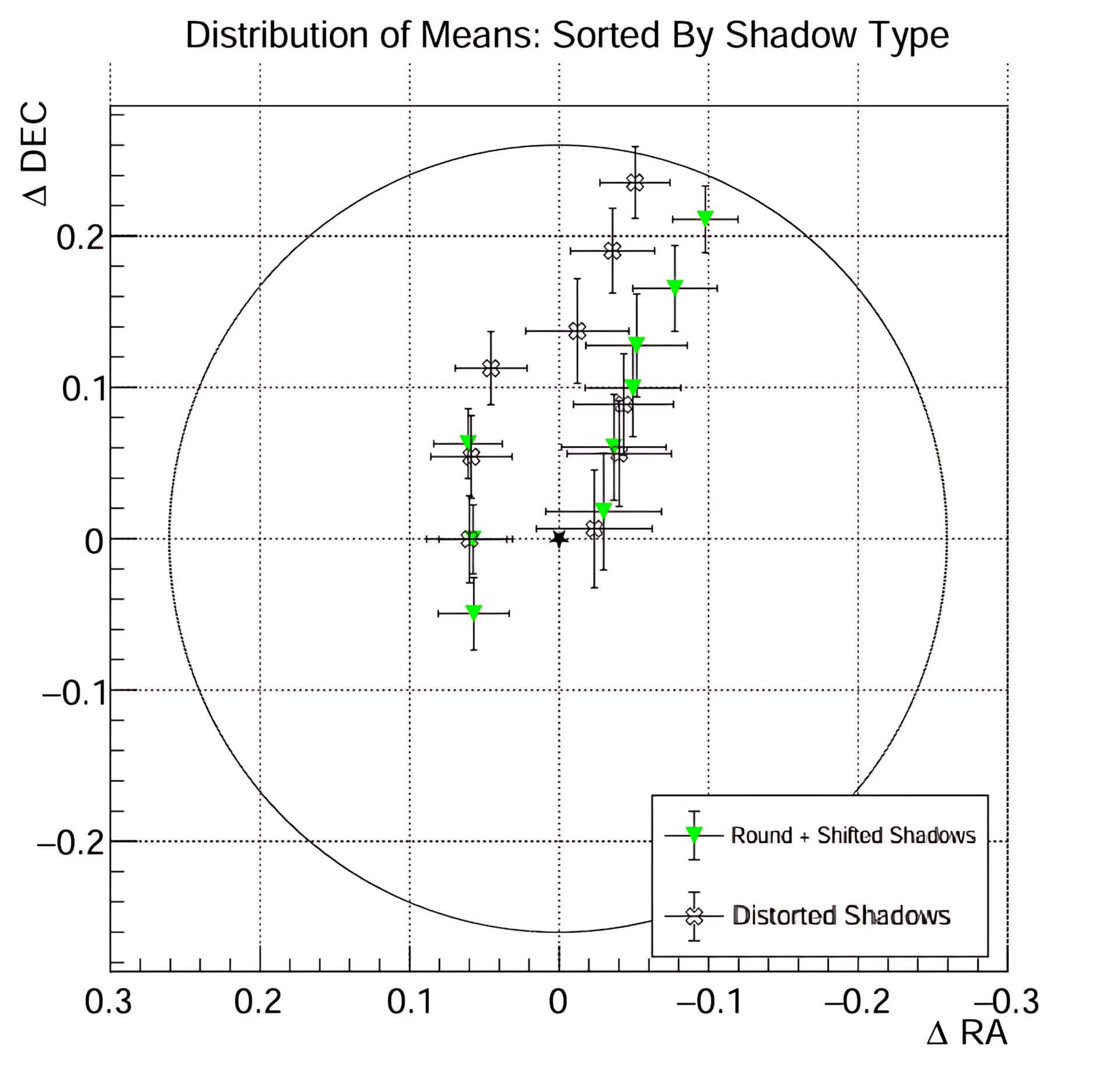}
        \label{fig:shadowTypeMeans2}
\qquad
        \centering
        \hspace{-0.6cm}
        \includegraphics[width=0.45\textwidth]{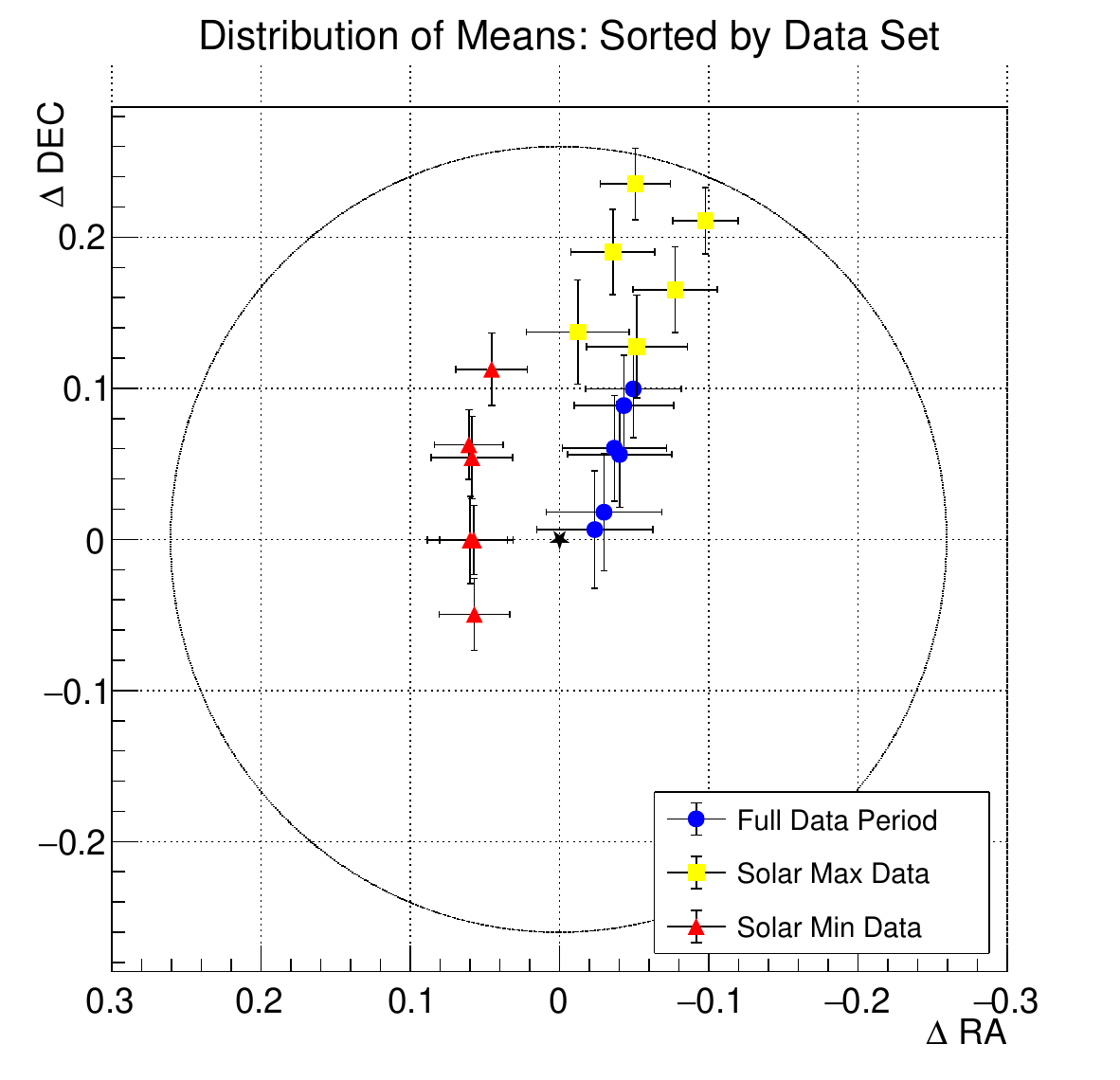}
        \label{fig:dataMeans2}
\caption[Scatter Plots of Final Data]{Scatter plot of the simulated shadow origin's $\Delta$RA and $\Delta$DEC means. $E_0$ was set to \unit[22.8]{TeV}. The same data is plotted in all 3 figures. The left figure categorizes the tests by polarity ratios. The center figure categorizes them by shadow template. The right figure categorizes them by the data sets. From these plots, each scenario is specified. The black circle in each plot marks the the boundary of the photosphere as viewed from Earth. All simulated shadow origins are within the photosphere.}
\label{fig:scatterB}
\end{figure}

Using this fit function, it was determined that a $\mu_{\alpha,w}$ value of 0 occurs at the primary cosmic ray energy cutoff of $E_{0} = 22.8 \pm \unit[1.2]{TeV} $. Note that it is 2-3 times higher than the accepted value of $\unit[10]{TeV}$. Using this $E_{0}$ value a more centered set of shadow distributions is produced using the same 18 scenarios from earlier. The new cutoff energy is the only difference in these simulations. The resulting full set of simulated origin trajectories is shown in Fig.~\ref{fig:scatterB} following the same format as Fig.~\ref{fig:scatterA}.

These results show consistent agreement of all shadow templates originating from the Sun. Templates from the full sampling period appears to be the most well reconstructed of the sets, followed by the solar minimum sets. This is expected, due to the full data set having a much more certain shadow location. The HMF tends to be much less dynamic at solar minimum \cite{Kallenrode:2004plasma} so it is not surprising that the solar max sets share the least agreement. The data seems to favor a positive polarity dominated HMF, especially for the solar max data. The distorted and simple round shadows both tend to deviate by similar amounts.

The determination of $E_{0} = 22.8 \pm \unit[1.2]{TeV}$ is contingent upon the Parker spiral model being an accurate representation of the HMF. This value is inconsistent with observation, indicating that there is more structure to the HMF than is accounted for in this base model. Significant shifts in the observed shadow location can be produced by slight alterations to the GCR energy spectrum.

\section{Extensions and Conclusions} 

\subsection{Zenith Corrected Energy Distribution}

\begin{figure} [h] 
\centering
    \centering
    \includegraphics[width=0.45\textwidth]{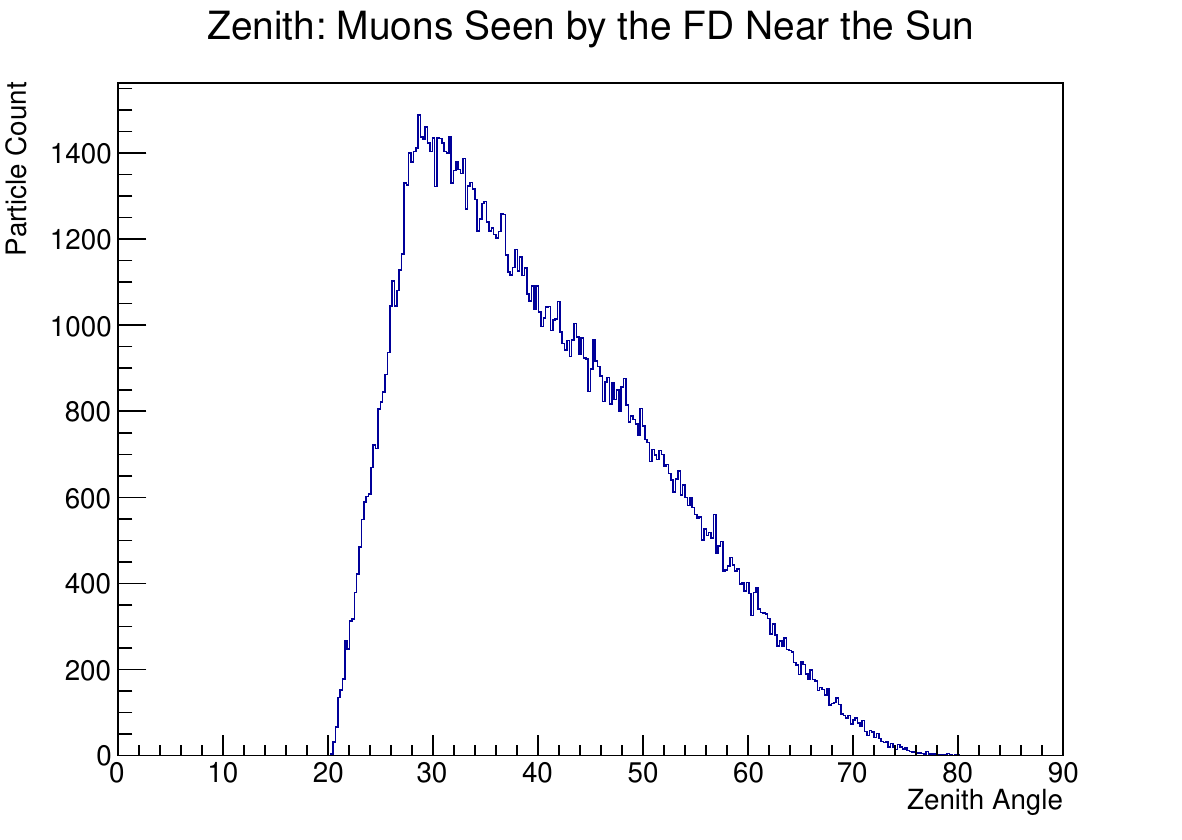}
    \label{fig:nearSunZenith}
\qquad
    \centering
    \includegraphics[width=0.45\textwidth]{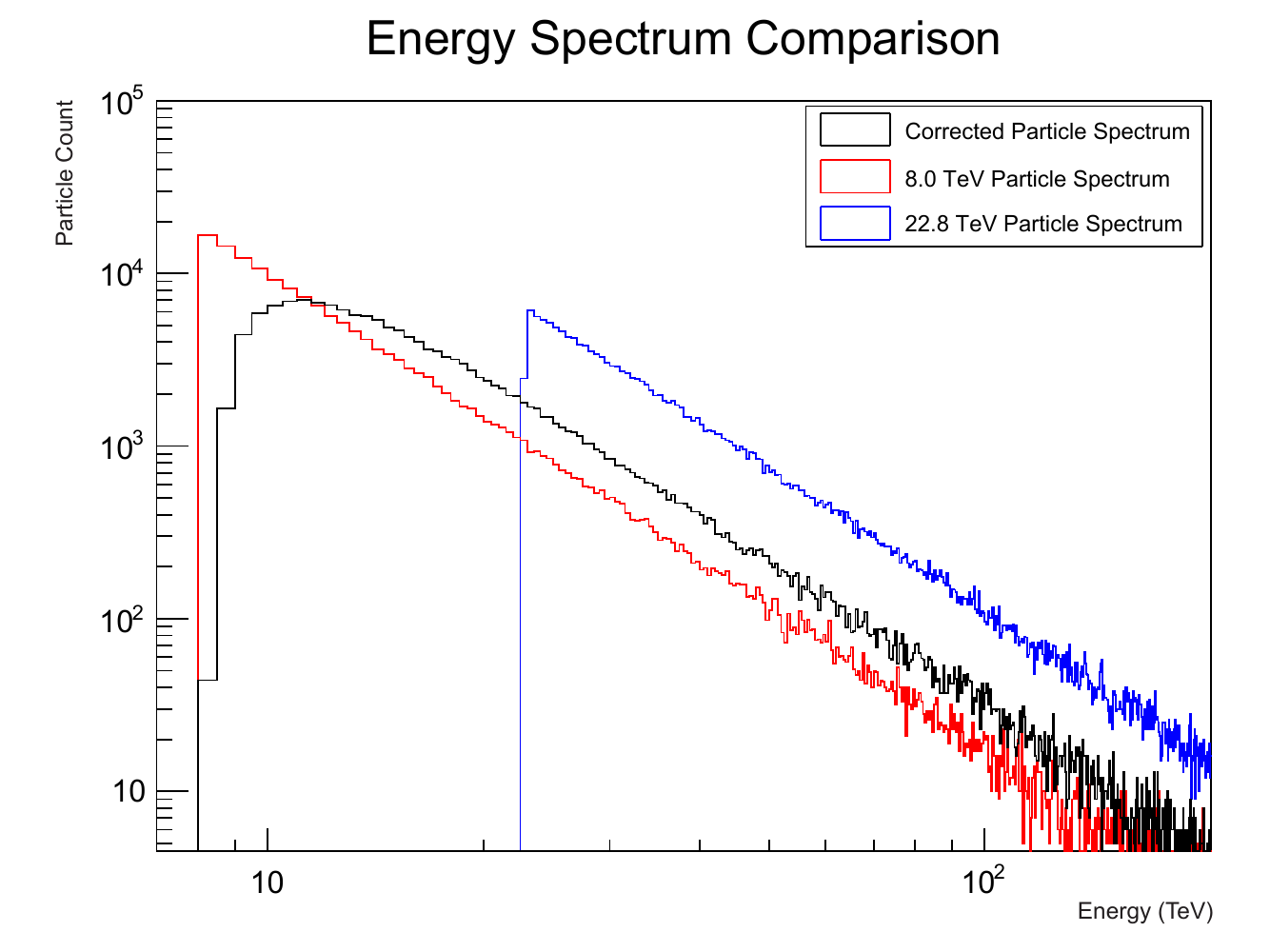}
    \label{fig:energyCorrectionComparison}
\caption[Energy Correction]{The left figure shows the zenith angles of the near Sun data particles (within $4^{\circ}$). The right figure compares the $\unit[8.0]{TeV}$ spectrum and the $E_{0} = \unit[22.8]{TeV}$ spectrum to the corrected spectrum drawn directly from the data. Energies are generated using minimum cutoff energies calculated from $\theta_{z}$ angles and randomly sampling a value from the CR power law distribution.}
\label{fig:energyCorrection}
\end{figure}

After the completion of the energy analysis detailed in Section 4.4, it was realized that because the minimum cutoff energy ($E_{0}$) depends on the distance the particles travel through rock then it must therefore also depend on the zenith angle of the incident particles. A cutoff of $\approx \unit[7.8]{TeV}$ corresponds to a particle penetrating from a zenith of $0^{\circ}$; $E_{0} = \frac{\unit[7.8]{TeV}}{\cos{\theta_{z}}}$ where $\theta_{z}$ is the zenith angle. Once a minimum cutoff energy is determined for any particular particle, a possible energy of the particle can be sampled using that minimum and the characteristic $E^{-2.75}$ distribution. The results are shown along with the initially tested $E_{0} = \unit[8.0]{TeV}$ spectrum and the $E_{0} = \unit[22.8]{TeV}$ spectrum in Fig. \ref{fig:energyCorrection}. From this plot it is clear that the actual spectrum behaves more like the $\unit[8.0]{TeV}$ spectrum than the $E_{0} = \unit[22.8]{TeV}$ spectrum. The corrected spectrum most closely mimics how a \unit[10]{TeV} spectrum would behave but again with a shorter peak and a higher volume tail. The weighted mean right ascension ($\mu_{\alpha,w}$) for the \unit[10.0]{TeV} minimum cutoff was $\mu_{\alpha,w} = -0.216 \pm 0.005^{\circ}$. Therefore, this correction alone would not be enough to explain the higher than expected $\Delta$RA shift and it would move the simulated shadow centers closer to their expected positions.

\subsection{Possible Additional Parameters}

The model tested in this study, the Parker spiral model, is a simple foundation upon which much more complex models can be built. There were a number of assumptions made to motivate its basic structure, but there are several aspects of it that are known to vary. The rotational period of the Sun ($\omega\textsubscript{\(\odot\)}$) is fairly stable as mentioned \cite{Kallenrode:2004plasma} but the solar wind velocity ($u\textsubscript{sowi}$) and photospheric magnetic field magnitude ($B\textsubscript{0}$) are variant and have significant effects on the overall field. Another set of simulations could be generated where $u\textsubscript{sowi}$ and $B\textsubscript{0}$ are allowed to vary freely. Varying $u\textsubscript{sowi}$ would produce a direct effect on the direction of the HMF field vectors more than anything else, given that it only affects the azimuthal component. Varying $B\textsubscript{0}$ would have an inverse effect to varying the energy. A stronger field makes the particles less rigid and a weaker field makes them more rigid. Even using the residual sims these variations would be difficult to test. Quicker less comprehensive methods such as the Krueger Plots in Fig.~\ref{fig:KruegerPlots} would allow dozens or hundreds of parameter combinations to be tested for general effectiveness before a larger scale more comprehensive survey is performed. Also, data from direct satellite measurements could allow for a more accurate representation of the polarity, velocity, and strength of the HMF in specific regions and times.

There are other aspects of the HMF that were simply ignored, because either their effects were deemed too rare to be relevant or their complexity is beyond the scope of this study. In any plasma, there exist regions of disturbances in the form of waves and shock fronts \cite{Kallenrode:2004plasma}. These disturbances exist in the Sun/Earth system most prominently in the form of the magnetosphere's bow shock and the HMF's corotating interaction regions. There is also the matter of the neutral current sheet which may deserve consideration. The neutral current sheet projects outward from the Sun's neutral line which is typically near the Sun's rotational axis but it sometimes develops a tilt. As the solar wind moves radially outward away from the photosphere, it is driven toward this neutral region which forms the neutral current sheet \cite{Kallenrode:2004plasma}. The spiral structure still forms as the Sun rotates and the result is a wave-like structure which increases in amplitude radially. It has been likened to a ballerina's skirt and therefore is known as the Ballerina Model of the HMF. It is significant because the neutral tilt line is known to move further from the solar equator during solar maximum \cite{Kallenrode:2004plasma}. This has a dramatic effect on the HMF structure and therefore it would be a good thing to include in a GCR motion simulation.

Any theoretical model would be extremely complex and therefore the subject of its own study. The solar wind velocity vector would take on a polar component which makes Gauss' Law no longer usable.  If this model could be implemented, it may show different collective polarity ratios than the ones determined in this study. It is certainly plausible that the large solar maximum $\Delta$DEC values observed in this study are due to a more ballerina-like structure during that period. It is also plausible that individual particles may experience multiple polarity changes throughout their journey from Sun to Earth. The wave-like structure of the Ballerina Model suggests each time a particle near the orbital plane crosses through the neutral current sheet, the polarity may change. This would cause the deflections to be partially self canceling, thus explaining why they appear to have deflected less than the base model.

\subsection{Conclusions}

The cosmic ray solar muon shadow is observable in the MINOS far detector muon data not only for the 13 year period from 2003-2016 but for much shorter time periods near solar max (2012-2014) and solar min (2007-2009). The shadows were located using a $\Lambda$ probability max-likelihood search technique with better than 6$\sigma$ certainty on the shadow's location for the whole time period and better than 3$\sigma$ certainty on both the solar max and solar min shadow locations. These shadows show distinctly different positions depending on the point in the solar cycle. A method for defining distorted shadow structures using the $\Lambda$ probability histograms has also been demonstrated.   

A functional magnetic field model and integration technique have been defined, implemented, and tested. A $4\textsuperscript{th}$ order Runge-Kutta scheme proves sufficient for modeling time reversible relativistic particle motions subject to the Lorentz force. A theoretical framework for superimposing the HMF and the GMF have been illustrated and implemented as well. Typical CGR motions have been observed, plotted, and discussed. The polarity of the HMF has shown to have a direct and measurable effect on the $\Delta$DEC deflection of galactic cosmic rays. The geomagnetic field and HMF together will typically shift GCRs to larger declination and right ascension positions, but the degree to which the overall distribution deflects depends greatly on HMF conditions and the primary cosmic ray energy spectrum. 

A primary cosmic ray energy distribution following a power law of $E^{-2.75}$ best models the observed shadows with a minimum cutoff energy of $E_{0} = 22.8 \pm \unit[1.2]{TeV}$ which is 2-3 times the experimentally determined value \cite{Adamson:ApP}. It has been directly shown that the position of the shadows is highly dependent on the energy spectrum of the corresponding GCR distribution. A number of untested factors remain relevant and perhaps responsible for the discrepancies between the shadow locations at different points in the solar cycle. These factors include neutral line tilt, impulses from shock fronts, and HMF polarity variations. This study finds the cosmic ray solar muon shadows observed solar minimum to be in best agreement with the dipolar model of the geomagnetic field and the Parker spiral model of the interplanetary magnetic field. The full data set and the solar max data set show clear indications of other factors at work, most likely some combination of the aforementioned effects. The general direction of deflection of CGRs due to the HMF field is consistent with the Parker spiral model as a base with more complex structures not yet modeled.

\end{document}